\documentclass[final,3p,times]{elsarticle}
\usepackage{graphicx}
\usepackage{caption}
\usepackage{subcaption}

\usepackage{amssymb}
\usepackage{amsmath}
\usepackage{multicol}

\journal{Journal Name}

\begin{document}

\begin{frontmatter}



\title{Properties of strange quark stars in $f(Q)$ gravity: A baryon number density-dependent perspective} 


\author[inst1]{Sourav Biswas}
\author[inst1]{Debadri Bhattacharjee}
\author[inst1]{Pradip Kumar Chattopadhyay} 
\address[inst1]{IUCAA Centre for Astronomy Research and Development (ICARD), Department of Physics, Cooch Behar Panchanan Barma University, Vivekananda Street, District: Cooch Behar,  Pin: 736101, West Bengal, India.}
\author[inst2]{Nawal H. Siddig} 
\address[inst2]{Department of Mathematical Sciences, College of Science, Princess Nourah bint Abdulrahman University, P.O. Box 84428, Riyadh 11671, Saudi Arabia.}
\author[inst3,inst4]{Euaggelos E. Zotos} 
\address[inst3]{Department of Physics, School of Science, Aristotle
	University of Thessaloniki, GR-541 24, Thessaloniki, Greece.}
\address[inst4]{S.M. Nikolskii Mathematical Institute of the Peoples'
	Friendship University of Russia (RUDN University), Moscow 117198, Russia.}
\author[inst2]{Laila Abdulaziz Al-Essa} 
\ead{souravcbpbu@gmail.com,debadriwork@gmail.com, pkc\_76@rediffmail.com, nhsiddig@pnu.edu.sa, evzotos@physics.auth.gr, Laalessa@pnu.edu.sa}

\begin{abstract}
We study anisotropic strange quark stars in $f(Q)$ gravity, incorporating a baryon-number-density-dependent bag function $B(n)$ into the MIT bag equation of state, $p=[\rho-4B(n)]/3$. Motivated by a Woods–Saxon parametrization, the density dependence of $B(n)$ provides a phenomenological description of the evolution of quark matter with increasing baryon density beyond the conventional constant-bag prescription. Adopting the Finch–Skea metric ansatz, we obtain exact solutions of the modified field equations for anisotropic matter composed of deconfined up, down and strange quarks. We investigate the thermodynamic stability through the energy per baryon, $E_B=\rho_0/n$, as a function of baryon density and identify the regimes corresponding to stable hadronic matter, metastable hybrid matter and absolutely stable strange quark matter. At sufficiently high densities, $E_B$ falls below the corresponding value for $^{56}$Fe, satisfying the stability criterion for strange quark matter. The resulting stability window, $57.55\leq B(n)\leq 91.55~\mathrm{MeV/fm^3}$, supports stable three-flavour quark matter. Solving the generalized Tolman-Oppenheimer-Volkoff equations, we find a maximum stellar mass of $\sim2.46,M_\odot$, demonstrating that the model can accommodate massive compact stars. The resulting configurations satisfy the relevant causality and energy condition constraints, while their stability is supported by the Herrera cracking criterion and the adiabatic index. These results demonstrate that the combined effects of nonmetricity and a density-dependent bag function can substantially enrich the phenomenology of strange quark stars and provide a viable framework for exploring their internal structure and stability in modified gravity.
\end{abstract}




\end{frontmatter}

\section{Introduction}\label{sec1}
Despite the great success of Newtonian gravity, it failed mostly under certain circumstances such as the precession of Mercury's orbit and the Michelson-Morley experiment \cite{Eisele} under strong gravitational effects. General Relativity (GR) made it possible to solve the precession of Mercury's orbit \cite{Wheeler}. GR is considered a basic theory to describe the physics of gravitational phenomena since then. Karl Schwarzschild \cite{Schwarzschild} provided the vacuum solution to Einstein's field equations (EFE), describing the gravitational field around a spherical, static, and compact object. 

Numerous groundbreaking discoveries have been made in this field, expanding our knowledge of the cosmos and the extreme conditions that exist within these enigmatic objects. While GR has been remarkably successful in explaining the solar system and large-scale cosmic structures, it still faces significant challenges. These include the nature of dark matter and dark energy \cite{Perlmutter,Riess,Riess1,Hirata,Dodelson,Cole}, the inflationary epoch of the early universe, and the difficulties in quantising gravity. Furthermore, GR shows inconsistencies in the strong gravitational field as evident from recent observations \cite{Hawkins,Spergel,Shekh}. Therefore, appropriate modifications in GR are required to explain such inconsistencies.

Compact objects are fascinating celestial bodies that bridge the gap between astrophysics, nuclear physics, and particle physics. While neutron stars are composed primarily of neutrons held together by gravity, strange stars are hypothesised to be made entirely of strange quark matter, bound by both strong interactions and gravitational forces.
It is widely accepted that GR might not be the final theory of gravity and could require modifications. One common approach is to modify the Lagrangian density by introducing functions of curvature, torsion, or non-metricity scalars. This leads to theories like $f(R)$ gravity \cite{Felice,Sotiriou}, $f(T)$ gravity \cite{Cai}-\cite{Casalino}, and $f(Q)$ gravity \cite{Beltran}-\cite{Capozziello} respectively. Recent studies have explored the similarities and differences between these theories in terms of symmetry breaking and degrees of freedom. In the existing literature, two such theories, mathematically and geometrically equivalent to GR, are often discussed: the teleparallel equivalent of GR, which utilises torsion \cite{Aldrovandi}-\cite{Linder}, and the symmetric teleparallel equivalent of GR, which uses non-metricity \cite{Nester}. While these theories are fundamentally the same as GR, their extensions differ significantly \cite{Altschul}.
$f(Q)$ formalism is a well-known modified theory to describe gravity that is an extension of symmetric teleparallel gravity. It replaces the geometric variable $Q$ (which represents non-metricity in symmetric teleparallel gravity) with a general function $f(Q)$ in the Lagrangian. Unlike general relativity, which uses the Levi-Civita connection, $f(Q)$ gravity utilises a connection that is both torsionless and curvatureless. This theory centres around non-metricity $(Q)$, which describes how it changes the length of a vector during parallel transportation. In the $f(Q)$ theory of gravity, Second-order equations are used, which are generally simpler to handle and lead to more manageable solutions. This is a significant advantage over $f(R)$ gravity, which has fourth-order field equations.  

Lazkoz et al. \cite{Lazkoz} introduced a crucial set of constraints on $f(Q)$ gravity, expressing the Lagrange density as a polynomial function of redshift. Utilising observational data from sources like Type Ia supernovae, cosmic microwave background distance measurements, Baryon Acoustic Oscillations (BAO), quasars, and gamma-ray bursts, researchers have successfully determined the range of applicability of these models. Recent studies \cite{Mandal1} have also explored $f(Q)$ gravity using observational data. Furthermore, $f(Q)$ gravity has been applied to investigate wormhole solutions \cite{Hasan}, its unique signatures \cite{Frusciante}, spherically symmetric configurations \cite{Lin}, and other aspects. Given its recent development, $f(Q)$ gravity has shown remarkable success in explaining the present state of the universe. This success motivates us to examine the stability of compact stars and develop physically viable stellar models within the framework of $f(Q)$ gravity. Consequently, investigating stellar structures within the context of $f(Q)$ gravity presents a novel and intriguing avenue of research. To solve the field equations, we have considered $e^{2\lambda}=1+kr^2$ as the $g_{tt}$ component proposed by Finch and Skea \cite{Finch}. Bhar et al. \cite{Piyali} proposed a new class of interior solutions of a $(2+1)-$ dimensional anisotropic star in Finch-Skea spacetime corresponding to the exterior BTZ black hole. A relativistic stellar model admitting a quadratic equation of state was proposed by Sharma and Ratanpal \cite{Sharma} by taking the Finch-Skea ansatz. Zubair et al. \cite{Zubair} studied the spherically symmetric compact star model with anisotropic matter distribution in the framework of $f(T)$ modified gravity by taking the Finch-Skea metric. Shamir et al. \cite{Shamir} explored the compact geometries by employing the Karmarkar condition with the charged anisotropic source of matter distribution by taking the Finch-Skea geometry. Bhar et al. \cite{Piyali1} discovered a new well-behaved charged anisotropic solution of Einstein-Maxwell's field equations under embedding class-I in the background of Finch-Skea geometry.

In our current investigation, we consider the compact star in presence of pressure anisotropy which arises when the pressure exerted radially $(p_{r})$ differs from the pressure exerted tangentially $(p_t)$, leading to unequal principal stresses. The anisotropy, quantified by the difference of tangential and radial pressures $(p_t-p_r)$, plays a crucial role in determining the internal forces. A negative anisotropy $(p_t-p_r<0)$ signifies an attractive force, drawing matter towards the centre of the celestial body. Conversely, a positive anisotropy $(p_t-p_r>0)$ induces a repulsive force, pushing matter away from the centre. At the centre of the celestial body, $p_t-p_r=0$. Ruderman \cite{Liebling} proposed that local anisotropy in compact stars could arise from a solid core. This idea was later formalised by Herrera and Santos \cite{Herrera1}. Dev and Gleiser \cite{Dev} demonstrated that transverse pressure is essential for spherical symmetry. Various physical phenomena can induce anisotropy within the interior of a compact star, such as exotic phase transitions, where extreme densities can trigger phase transitions into exotic states of matter \cite{Sokolov}. Superconductivity, where the presence of Type II superconductors within the star can introduce anisotropy \cite{Jones}. The condensation of pions, a type of subatomic particle, can also lead to anisotropic pressure \cite{Sawyer}. Type 3A Superfluidity, which is an exotic state of matter, can contribute to anisotropic pressure \cite{Kippenhahn}. Intense magnetic fields can also induce anisotropic stress-energy tensors \cite{ Ruderman}. The presence of Scalar fields in Boson Stars can lead to anisotropic pressure distributions \cite{Weber}.

The transition of phase from confined hadronic matter to the deconfined quark matter is supposed to occur when the pressure from the dense matter overcomes the pressure of the bag wall containing the quarks, causing the bag constant to effectively zero as the two vacuum states become indistinguishable. Therefore, for a more physically accurate description, the bag constant ($B$) should not be treated as a fixed value, but rather as a quantity that varies with the baryon number density ($n$) \cite{Burgio,Reinhardt}. As the baryon number density $(n)$ increases, the separation between the two vacuum states theoretically disappears. The primary motivation for adopting a baryon number density dependent bag parameter $B(n)$, parameterised via a Wood-Saxon type functional form \cite{Woods} within the framework of linear $f(Q)$ action, to derive a more physically realistic Equation of State (EoS). This modified EoS is essential for precise evaluation of gross properties of strange stars (SS), including their mass-radius relationship, stability criteria, adherence to causality conditions and fulfilment of relevant energy conditions. 

The MIT bag model provides a phenomenological description of deconfined quark matter, in which the bag parameter represents the difference between perturbative quark vacuum and the physical vacuum. However, treating the bag parameter as a constant may not adequately capture the density dependence of the effective confinement properties of strongly interacting matter. A density-dependent bag function $B(n)$ therefore provides a more flexible description of quark matter over the range of densities encountered inside compact stars. Furthermore, realistic compact star matter need not remain perfectly isotropic. Various microscopic and macroscopic mechanisms can generate a difference between the radial and tangential pressures, $p_r$ and $p_t$. Incorporating pressure anisotropy consequently allows us to investigate how deviations from the idealized isotropic configuration influence the equilibrium and stability of strange stars. On gravitational side, $f(Q)$ gravity modifies the geometric sector of Einstein's theory through a function of the non-metricity scalar $Q$. Thus, the matter-sector effects associated with density-dependent quark confinement and pressure anisotropy can be investigated together with possible modifications of the gravitational interaction. The combined framework consequently provides an opportunity to examine the interplay between microphysical properties of strange quark matter, anisotropic stellar structure and modified gravity. 
In the present study, these physical aspects are considered within a unified framework to investigate the interplay between the microphysics of strange quark matter, pressure anisotropy and modified gravitational dynamics. The MIT bag model offers a phenomenological description of deconfined quark matter, while allowing the bag parameter to depend on the baryon number density, $B=B(n)$, accounts for the possible variation of the effective confinement energy with the density of matter inside the star. This consideration is particularly relevant because the density varies considerably from the stellar center toward the surface, making a constant bag parameter a restrictive assumption. In addition, the assumption of isotropic pressure $p_r=p_t$, is an idealization that may not adequately represent the internal physics of compact objects. Anisotropic stresses, characterized by $p_t\neq p_r$, can modify the hydrostatic balance and consequently influence the mass, radius and stability of the stellar configuration. At the same time, $f(Q)$ gravity provides a modified description of the gravitational sector in which the gravitational dynamics are governed by a general function of the non-metricity scalar. Therefore, considering $B(n)$, pressure anisotropy and $f(Q)$ gravity simultaneously enables us to examine how modifications in the matter sector and gravitational sector interact in determining the structure of strange compact stars. The specific physical question addressed in this work is how the density-dependent confinement properties and anisotropic stresses of strange quark matter alter the equilibrium and stability of compact stars when the underlying gravitational dynamics depart from those of General Relativity. 

Motivated by these considerations, the present work aims to construct and systematically analyze anisotropic strange compact star configurations within $f(Q)$ gravity using the MIT bag model with a density-dependent bag parameter. We investigate how the model parameters governing the gravitational sector and density dependence of the bag function affect the stellar mass-radius relation, internal pressures and density distributions, equilibrium configuration, causality, energy conditions and stability properties. This framework allows us to assess whether physically viable strange star configurations can be obtained when realistic density-dependent quark-matter properties and pressure anisotropy are incorporated into modified gravity. 

This manuscript is arranged as follows: In section \ref{sec2}, we have discussed the baryon number density $(n)$ dependent bag parameter $B(n)$ and the expression of energy per baryon $(E_B=\rho_0/n)$. $E_B$ vs $n$ graph is plotted and its nature is analysed to evaluate the permitted values of baryon number density $(n)$. Then, the modified values of bag parameters are calculated from the values of baryon number density and the model, and its parameters are analysed for different values of $B(n)$. In section \ref{sec3}, we presented the elementary theory of $f(Q)$ modified gravity. In section \ref{sec4}, we derived the non-metricity scalar $Q$ for the model and considering the anisotropic perfect fluid distribution, we derived the EFE's. In section \ref{sec5}, we choose the Finch-Skea metric potential to construct the proposed model and derive the expression of energy density ($\rho$), pressure profiles ($p_r$ and $p_t$) and the expression of anisotropy ($\Delta$). Section \ref{sec6} deals with the boundary condition in the framework of $f(Q)$ gravity. Section \ref{sec7} represents the maximum mass and radius in the model for different values of baryon number density $(n)$. In section \ref{sec8}, we have discussed the physical properties along with the causality and energy conditions for the proposed model. All the stability conditions are analysed in section \ref{sec9}. In section \ref{sec10}, we derive the radius of various compact stars and present the result in tabulated form. Section \ref{sec11} provides a concise summary of the main findings.

\section{Effect of Baryon number density ($n$) on Bag parameter} \label{sec2}
Many compact stars are believed to be composed of quarks \cite{Itoh,Annala}. While the behaviour of the outer layer of these stars, which is primarily made of hadrons, is relatively well-understood, the properties of the quark matter core remain uncertain. A crucial aspect of understanding these stars lies in accurately describing the Equation of State (EoS) that governs their internal pressure and energy density, especially at densities far exceeding those found in atomic nuclei. At such extreme conditions, the presence of particles like hyperons, delta resonances, and mesons is anticipated, potentially leading to a phase transition from ordinary nuclear matter to a state known as quark-gluon plasma (QGP) \cite{Witten,Baym,Glendenning}. The MIT bag model provides a framework for describing this deconfined quark phase \cite{Kapusta,Chodos}.\\
Experiments conducted at CERN have demonstrated that the QGP created in high-energy heavy-ion collisions exhibits a low baryon number density and high temperature. This is in contrast to the expected conditions within neutron stars, where the baryon density is high, and the temperature is relatively low. The MIT bag model, while acknowledged to have some thermodynamic limitations, predicts a constant energy density for the deconfined quark phase \cite{Cleymans}. By adjusting the model's parameters to align with the experimental observations from CERN, this constant energy density allows for a simplified analysis by considering only zero temperature. Following the approach of Burgio et al. \cite{Burgio}, the boundary between the hadronic and quark phases is determined by the point where their respective energy densities become equal. Experimental results from CERN-SPS data suggest that this transition occurs at an energy density of approximately $1~GeV/fm^3$. Our model assumes that the transition of phase between hadronic and quark matter is primarily driven by energy density. The QCD phase diagram in the plane of temperature-chemical potential is primarily influenced by the value of the bag parameter $B$ in the context of the MIT bag model. To characterize the phase transition observed in the CERN experiment and incorporate it into stellar modeling using the quark core hypothesis, we employ an extreme Wood Saxon-like profile \cite{Woods} for the baryon number-dependent bag parameter $B(n)$, which is expressed as:
\begin{equation}
	B(n)=B_{as}+(B_0 - B_{as})e^{-\beta_n (\frac{n}{n_0})^2}, \label{eq1a}
\end{equation}
where, $B_{as}$ is the finite value of $B$ at asymptotic densities, $B_0=B(\rho =0)$, $\beta_n$ is the free parameter resulting from the transition density of $1.1~GeV/fm^3$ at which nuclear matter transform into quark phase, $n$ is the baryon number density and $n_0$ represents the baryon number density of ordinary nuclear matter. Following Relativistic Mean Field theory (RMF) \cite{Ghosh,Sahu}, we consider $B_{as}=38~MeV/fm^3$, $B_0=200~MeV/fm^3$, $\beta_n=0.14$ and $n_0=0.17~fm^{-3}$.\\
The energy per baryon $(E_B=\rho_0/n)$ of the most stable nucleus $^{56}Fe$ is about $930.4~MeV$ \cite{Madson}. The Bodmer-Witten hypothesis \cite{Witten,Bodmer} provides a criterion for determining whether strange quark matter (SQM) can be absolutely stable. According to this criterion, the energy per baryon of SQM, evaluated at zero external pressure, should be lower than that of $^{56}Fe$. Thus, SOQ is considered stable when $(E_B=\rho_0/n) < 930.4~MeV$, where $\rho_0$ and $n$ denotes the energy density and baryon number density of quarks respectively. Now, according to thermodynamics, the pressure can be related to the variation of $(\rho_0/n)$ with respect to the baryon number density $(n)$ through $p=\rho_0^2 \frac{d}{d\rho_0}(\frac{\rho_0}{n})$. This relation implies that minimum value of $(\rho_0/n)$ should correspond to a state with zero external pressure. However, this assumption was found to be incompatible in a reformulated quark mass dependent model \cite{Benvenuto,Lugones1}, where an inconsistency arises between the behavior of $(E_B=\rho_0/n)$ and the pressure $(p)$. \\
This issue was later addressed in a study of SQM based on a density-dependent quark mass model \cite{Lugones2} in which the inconsistency was resolved by formulating the problem within the canonical ensemble and it was shown that the the energy per baryon attains its minimum at zero pressure. In this approach, the EoS is derived using the Euler thermodynamic relation;
\begin{equation}
	\rho_0=-p+\sum_{i}^{} \mu_i n_i, \label{eq1}
\end{equation}
where $\rho_0$ and $p$ denotes the energy density and the pressure respectively, $\mu_i$ and $n_i$ represent the chemical potential and number density of the $i^{th}$ quark particle respectively. From the work of Debadri \cite{Deba}, we consider the expression of pressure defined as;
\begin{equation}
	p=\frac{3\mu^4}{4\pi^2}-B(n)+n\frac{\delta B(n)}{\delta n}. \label{eq2}
\end{equation}
Here, $\sum_{f}^{} m_f=0$ is considered. 
The quark number density can be formulated explicitly for the density dependent approach as follows;
\begin{equation}
	n=\frac{1}{3} \sum_{f}^{} n_f = -\frac{1}{3} \sum_{f}^{} \frac{\delta \Omega_f}{\delta \mu_f} = \sum_{f}^{} \frac{k_f^3}{\pi^2}.
\end{equation}
where, $k_f=\sqrt{\mu_f^2-m_f^2}$ represents the Fermi momentum of the free quark matter.
\begin{eqnarray}
	\rho_0&=&-p+\mu_i n_i \nonumber \\
	&=& -\frac{3\mu^4}{4\pi^2}+B(n)-n\frac{\delta B(n)}{\delta n}+ \frac{3\mu^4}{\pi^2} \nonumber \\
	&=& \frac{9\mu^4}{4\pi^2}+B(n)-n\frac{\delta B(n)}{\delta n}, \label{eq}
\end{eqnarray}	
\begin{figure}[h]
	\centering
	\includegraphics[width=10cm]{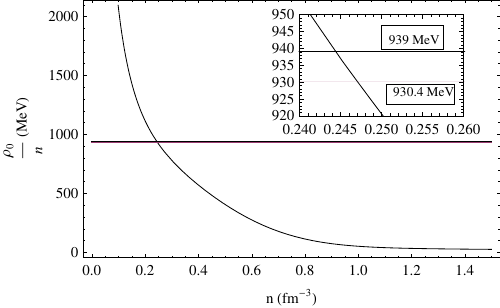}
	\caption{Variation of energy per baryon $(E_B=\rho_0/n)$ with baryon number density $(n)$}.
	\label{figa1}
\end{figure}
\begin{table}[!h]
	\centering
	\caption{\label{tab1} Constraint on baryon number density ($n$).}
	\vspace{0.5cm}
	\begin{tabular}{|ccc|}
		\hline
		unstable region & meta-stable region & stable region \\ 
		$(\rho_0/n)>~939~MeV$ & $930.4 <(\rho_0/n)< 939~MeV$ & $(\rho_0/n) < 930.4~MeV $  \\
		\hline 
		\vspace{0.1cm}
		$n < 0.244 $ & $0.244 < n <0.247$ & $n > 0.247$ \\
		\hline
	\end{tabular}
\end{table} 
where $B(n)$ is taken from Eq.~(\ref{eq1a}).
The value of energy per baryon in the case of the most stable nucleus $^{56}Fe$ is $930.4~MeV$. Considering a system of three flavour quarks composed of $u$, $d$ and $s$, stable quark matter is possible when $(E_B=\rho_0/n)< 930.4~MeV$ at zero external pressure. The range $930.4~MeV < E_B=\rho_0/n < 939~MeV$ is termed as meta-stability \cite{Backes}, and above the value $(E_B=\rho_0/n) > 939~MeV$, the quark matter becomes unstable. However, in the presence of pressure, inside the star, the energy per baryon may be greater than $930~MeV$ and strange matter may be stable in this situation too, according to Bodmer \cite{Bodmer} and Witten \cite{Witten} hypothesis. As we are dealing with the bare strange star modeling, we consider the value of binding energy per baryon below the value $930.4~MeV$ at the surface of the star.  Therefore, when $(E_B=\rho_0/n) > 939~MeV$, a stable SQM is forbidden and the nature of the interior matter is expected to be hadronic accordingly. The stellar configuration, in this case, may be termed a neutron star.\\
In Fig.~\ref{figa1}, the variation of energy per baryon $(E_B=\rho_0/n)$ with the baryon number density $(n)$ is shown. From Fig.~\ref{figa1}, it is noted that when $n<0.244~fm^{-3}$, the matter is found to be unstable in view of the quark phase but may be stable with respect to hadronic phase as in this case, $(E_B=\rho_0/n)>939~MeV$ for $n<0.244~fm^{-3}$. On the other hand, a possible change from hadronic phase to deconfined quark phase may start when $n=0.244~fm^{-3}$. In this situation interior matter switches to the mixed phase composed of hadrons and quarks, termed the meta-stable state. Meta-stability exists only when $n$ lies within the range $0.244\leq n\leq 0.247~fm^{-3}$. From Fig.~\ref{figa1}, it is also evident that with the increase of the baryon number density $(n)$, the energy per baryon $(E_B=\rho_0/n)$ decreases. For $n>0.247~fm^{-3}$, the value of $(E_B=\rho_0/n)<930.4~MeV$. It is evident from Fig.~\ref{figa1} that, for $n>0.247~fm^{-3}$, all hadrons may be converted into quarks and the transformed matter becomes stable. We consider arbitrary values of $n$ and substitute them into Eq.~(\ref{eq1a}) to obtain the corresponding values of $B(n)$ which are tabulated in Table~\ref{tab:1a}. The tabulated values are then used to study the physical analysis of a stellar configuration composed of stable quark matter.
\begin{figure}[h!]
	\centering
	\includegraphics[width=10cm]{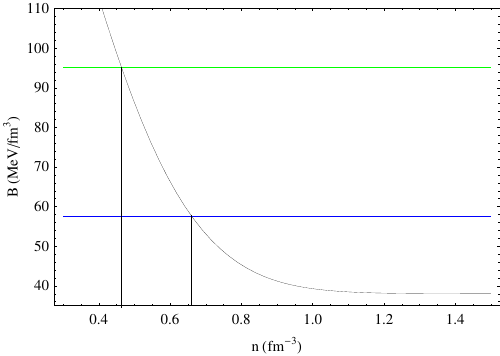}
	\caption{Variation of Bag parameter $B(n)$ with baryon number density $(n)$.}
	\label{figb}
\end{figure}
\begin{table}[h!]
	\centering
	\caption{Value's of $B(n)$'s corresponding to baryon number density$(n)$}
	\label{tab:1a}
	\begin{tabular}{|c|c|}
		\hline
		$n~(fm^{-3})$ &  $B(n)~(MeV/fm^3)$  \\ \hline
		${\bf 0.463}$ & ${\bf 95.11}$ \\
		$0.478$ & $91.55$ \\
		$0.563$ & $72.90$ \\ 
		$0.578$ & $70.11$ \\ 
		$0.608$ & $65.02$ \\ 
		${\bf 0.660}$ & ${\bf 57.55}$ \\
		$0.871$ & $42.10$ \\
		$1.110$ & $38.41$ \\
	    \hline
	\end{tabular}
\end{table}
\section{Elementary theory of f(Q) gravity}\label{sec3}
Considering the framework of symmetric tele-parallel $f(Q)$ gravity the gravitational action integral is expressed as \cite{Heisenberg,Zhao}:
\begin{equation}
	s=\int \sqrt{-g}d^4x\bigg[\frac{1}{2}f(Q)+\lambda_l^{kij} R_{kij} ^l +\tau_k^{ij} T_{ij}^k +L_m \bigg], \label{eq1}
\end{equation}
where, the determinant of the fundamental metric tensor is denoted by $g_{ij} = |g_{ij}|$, $f(Q)$ is a function of $Q$, known as non-metricity, $\lambda _l ^{kij}$ and $\tau _{ij} ^k$ are the Riemann tensor and torsion tensor respectively and $L_m$  defines the Lagrangian density of matter. Using affine connections ($\Gamma _{ij} ^k$), the non-metricity is expressed as:
\begin{equation}
	Q_{kij} =\nabla_k g_{ij} = \delta_k g_{ij} -\Gamma'_{ij} g_{ij} -\Gamma'_{ik} g_{jl}, \label{eq2} 
\end{equation}
where $\Delta_k$ is termed the covariant derivative. The affine connection is expressed by the addition of three components in the following way:
\begin{equation}
	\Gamma_{ij} ^k =\epsilon_{ij}^k + K_{ij}^k +L_{ij}^k. \label{eq3}
\end{equation}
Here, $\epsilon_{ij}^k$ defines the Levi-Civita connection, which, in terms of the fundamental tensor $g_{ij}$, is expressed as:
\begin{equation}
	\epsilon_{ij}^k =\frac{1}{2}g_{kl}(\partial_i g_{lj}+\partial_j g_{il}-\partial_l g_{ij}), \label{eq4}
\end{equation}
$K_{ij}^k$ indicates the contorsion and is expressed as:
\begin{equation}
	K_{ij}^k =\frac{1}{2} T_{ij}^k +T_{ikj}. \label{eq5}
\end{equation}
The anti-symmetric part related to the affine connection is physically represented by the contorsion $(K^{k}_{ij})$ in the STEGR, which is expressed as: $T_{ij}^k =2\Gamma_{[ij]}^k =\Gamma_{ij}^k -\Gamma_{ji}^k$ .\\
In Eq~(\ref{eq3}), $L_{ij}^k$ represents the deformation and is expressed as:
\begin{equation}
	L_{ij}^k =\frac{1}{2}Q_{ij}^k +Q_{(i k j)}. \label{eq6}
\end{equation}
The super potential corresponding to the non-metricity is described as:
\begin{equation}
	P^{kij} =-\frac{1}{4} Q^{kij} +\frac{1}{2} Q^{(ij)k} +\frac{1}{4}(Q^k-\tilde{Q}^k)g^{ij}-\frac{1}{4}\delta^{k(iQj)}, \label{eq7}
\end{equation}
where, $Q^k$ and $\tilde{Q}^k$ define the independent traces of $Q_{kij}$ as:
\begin{eqnarray}
	Q_k \equiv Q_{ki}^i, \nonumber \\
	\tilde{Q}^k =Q_i ^{ki}. \label{eq8}
\end{eqnarray}
Finally, the non-metricity scalar is expressed as:
\begin{equation}
	Q=-g^{ij}\big(L_{lj}^k L_{ik}^l - L_{il}^l L_{ij}^k \bigg)=Q_{kij}P^{kij}. \label{eq9}
\end{equation}
The affine connection can be characterised by a set of tractable functions without the effect of torsion and curvature, and described as:
\begin{equation}
	\Gamma_{ij}^k =\bigg(\frac{\partial x^k}{\partial \chi^l}\bigg) \partial _i \partial _j  \chi ^l, \label{eq10}
\end{equation}
where $\chi^l$ denotes arbitrary spacetime position functions, in the light of a general coordinate transformation.
Significantly, we have the flexibility to choose the coordinate in such a form $\chi ^l=\chi^l(x^i)$ always.
As the study of Jimenez et al.\cite{Zhao}, the generality of the affine connection, i.e., $\Gamma_{ij}^k =0$, is termed the property of coincident gauge. Hence, in the standard GR, the covariant derivatives reduce to ordinary derivatives in the coincident gauge and as a result, the non-metricity formalism of Eq.~(\ref{eq3}) can be simplified as:
\begin{equation}
	Q_{kij}=\partial_k g_{ij} \label{eq11}
\end{equation}
Now, by varying the Einstein-Hilbert action as written in Eq.~(\ref{eq1}), we obtain the tractable set of gravitational field equations in terms of $g_{ij}$ as:
\begin{equation}
	\frac{2}{\sqrt{-g}}\nabla_k \bigg(\sqrt{-g}f_Q P_{ij}^k \bigg) +\frac{1}{2}g_{ij} f+ f_Q \bigg(P_{ikl}Q_j^{kl}-2Q_{kli}P_j^{kl} \bigg)=-T_{ij}, \label{eq12}
\end{equation}
where, $f_Q$ denotes partial derivative of function scalar $f$ with respect to non-metricity $Q$. $T_{ij}$ represents the energy-momentum tensor of matter distribution and can be described in a generalised way in the following form:
\begin{equation}
	T_{ij}=-\frac{2}{\sqrt{-g}}\frac{\delta(\sqrt{-g}L_m)}{\delta g^{ij}}. \label{eq13}
\end{equation}
Additionally, the variation of action shown in Eq.~(\ref{eq1}) in terms of affine connection yields,
\begin{equation}
	\nabla_i \nabla_j \bigg(\sqrt{-g}f_Q P_{ij}^k-\frac{1}{2}\frac{\delta L_m}{\delta \Gamma_{ij}^k}\bigg)=0, \label{eq14}
\end{equation}
where the second term of the above equation defines the tensor density of the hyper momentum. If we substitute the relation, $\nabla_i\nabla_j (H_{ij}^k)=0$ in Eq.~(\ref{eq14}), we get $\nabla_i\nabla_j (\sqrt{-g}f_Q P_{ij}^k)=0$.

\section{Einstein field equations considering $f(Q)$ gravity} \label{sec4}
To construct the model, we consider that in the interior of a star, the space-time is spherically symmetric and static, which is characterised by the following line element:
\begin{equation}
	ds^2=-e^{2\nu(r)} dt^2+e^{2\lambda(r)}dr^2+r^2(d\theta^2 + sin^2\theta d\phi^2), \label{eq15}
\end{equation}
where $\lambda(r)$ and $\nu(r)$ are the metric potentials and are functions of the radial coordinate only. The Einstein field equation (EFE) is written as, $G^{\mu \nu}=-8 \pi G T^{\mu \nu}$, where $G^{\mu \nu}$ is Einstein's tensor, $G$ is known as the Newtonian gravitational constant. $T^{\mu \nu}$ is the stress-energy tensor. We employ a system of units in which $c=1, ~ G=1$.  \\
Using Eq.~(\ref{eq15}) in Eq.~(\ref{eq9}), we derive the expression of the non-metricity scalar given below:
\begin{equation}
	Q=-\frac{2e^{-2\lambda(r)}}{r}\bigg(2\nu'(r)+\frac{1}{r}\bigg), \label{eq16}
\end{equation}
where primes denotes the derivative with respect to $r$. Now, the anisotropic perfect fluid distribution is characterised by the following energy-momentum tensor: 
\begin{equation}
	T_{ij}=(\rho+p_t)u_i u_j +p_t g_{ij} + (p_r -p_t)v_i v_j, \label{eq17}
\end{equation}  
where $u_i$ refers to the fluid's four velocity and $v_i$ represents the space-like unit vector pointing in the radial direction. $u_i$ and $v_i$ obey the relations $u^i u_i =-1$ and $v^i v_i =1$ respectively. The terms $\rho$, $p_r $ and $ p_t$ in Eq.~(\ref{eq17}) represent the energy density, radial and tangential pressures, respectively.\\
Following the equation of motion given in Eq.~(\ref{eq12}) along with the anisotropic perfect fluid distribution of Eq.~(\ref{eq17}), we obtain a tractable set of the non-zero components of the EFE as follows:
\begin{eqnarray}
	\frac{f(Q)}{2}-f_Q\bigg[Q+\frac{1}{r^2}+\frac{2e^{-2\lambda}}{r}(\nu'+\lambda')\bigg]&=&8\pi\rho, \label{eq18} \\
	-\frac{f(Q)}{2}+f_Q\bigg[Q+\frac{1}{r^2}\bigg]&=& 8\pi p_r, \label{eq19} \\
	-\frac{f(Q)}{2}+f_Q\bigg[\frac{Q}{2}-e^{-2\lambda}{\nu''+2(\frac{\nu'}{2}+\frac{1}{2r})(\nu'-\lambda')}\bigg]&=&8\pi p_t, \label{eq20} \\
	\frac{cot\theta}{2}Q'f_{QQ}&=&0. \label{eq21} 
\end{eqnarray}
Now, one can obtain a linear form of f(Q) action by utilizing Eq.~(\ref{eq21}) and can be written as:
\begin{equation}
	f(Q)=\alpha_0+\alpha_1 Q, \label{eq22}
\end{equation}
where, the dimensions of $\alpha_0$ is $Km^{-2}$ and $\alpha _1$ is dimensionless.\\
Now, using Eq.~(\ref{eq22}), the field equations given in Eqs.~(\ref{eq18}), (\ref{eq19}) and (\ref{eq20}) in the modified $f(Q)$ theory of gravity can be reduced to an exact set of equations given below:
\begin{eqnarray}
	\frac{1}{2r^2}\bigg[r^2\alpha_0-2\alpha_1e^{-2\lambda}(2r\lambda'-1)-2\alpha_1 \bigg]&=&8\pi\rho, \label{eq23} \\
	\frac{1}{2r^2}\bigg[-r^2\alpha_0 -2\alpha_1e^{-2\lambda}(2r\nu'+1)+2\alpha_1 \bigg]&=&8\pi p_r, \label{eq24} \\
	\frac{e^{-2\lambda}}{2r}\bigg[-r\alpha_0e^{2\lambda}-2r\alpha_1\nu''-2\alpha_1(r\nu'+1)(\nu'-\lambda')\bigg]&=&8\pi p_t. \label{eq25} 
\end{eqnarray}
It is noted that $\rho,~ p_r$ and $p_t$ depends on the parameters $\alpha_0$ and $\alpha_1$.

\section{The proposed model using the Finch-Skea metric potential } \label{sec5}
In Eqs.~(\ref{eq23})-(\ref{eq25}), there are five unknown quantities, namely $\lambda,~\nu,~\rho,~p_r$ and $p_t$. To solve the system of equations, we have to fix any two of them. Among different ways to do the same, we choose the $g_{rr}$ component of the metric ansatz as proposed by Finch-Skea \cite{Finch},
\begin{equation}
	e^{2\lambda}=1+kr^2, \label{eq26}
\end{equation}
where $k$ is a constant which explicitly depends on the modified boundary conditions and has the dimension of $Km^{-2}$.
Substituting  Eq.~(\ref{eq26}), in  Eq.~(\ref{eq23}), we obtain,
\begin{eqnarray}
	\rho &=& \frac{\alpha_0+2kr^2\alpha_0+k^2r^4\alpha_0-6k\alpha_1-2k^2r^2\alpha_1}{16\pi (1+kr^2)^2}, \label{eq27}
\end{eqnarray}
Now, we consider the well-known EoS for strange quark matter in the MIT Bag model, expressed as:
\begin{equation}
	p_r=\frac{1}{3}(\rho-4B(n)), \label{eq28}
\end{equation} 
where we have considered the baryon number density dependent bag parameter as $B(n)$ having the form described in Eq.~(\ref{eq1a}). Plugging Eq.~(\ref{eq27}) in Eq.~(\ref{eq28}), the expression of radial pressure can be written as:
\begin{eqnarray}
	p_r &=& \frac{1}{3}\bigg(\frac{\alpha_0+2kr^2\alpha_0+k^2r^4\alpha_0-6k\alpha_1-2k^2r^2\alpha_1}{16\pi (1+kr^2)^2}-4B(n) \bigg). \label{eq29} 
\end{eqnarray}
Substituting Eqs.~(\ref{eq26}) and (\ref{eq29}), in Eq.~(\ref{eq24}), we obtain, 
\begin{equation}
	\nu=\frac{1}{3}\bigg(kr^2+\frac{(1+kr^2)^2(16B(n)\pi-\alpha_0)}{4k\alpha_1}+\frac{1}{2}\log(1+kr^2)\bigg). \label{eq30}
\end{equation}
Using Eqs.~(\ref{eq26}) and (\ref{eq30}), in Eq.~(\ref{eq25}), the expression of tangential pressure in this model will be:
\begin{eqnarray}
	p_t &=& \frac{1}{144\pi(1+kr^2)^3\alpha_1}(-512B(n)^2\pi^2r^2(1+kr^2)^4-2k^4r^{10}\alpha_0^2 +3\alpha_1(\alpha_0-6k\alpha_1) \nonumber \\ &&+8k^3r^8\alpha_0(-\alpha_0+k\alpha_1)+kr^4(-8\alpha_0^2+53k\alpha_0\alpha_1-36k^2\alpha_1^2)+r^2(-2\alpha_0^2+27k\alpha_0\alpha_1-30k^2\alpha_1^2) \nonumber \\
	&&+k^2r^6(-12\alpha_0^2+37k\alpha_0\alpha_1-8k^2\alpha_1^2) \nonumber \\ &&+32B(n)\pi(1+kr^2)^2(2k^2r^6\alpha_0-6\alpha_1+r^2(2\alpha_0-15k\alpha_1)+4kr^4(\alpha_0-k\alpha_1))). \label{eq31}
\end{eqnarray}
The difference in radial and tangential pressure, termed as anisotropy parameter ($\Delta$), is expressed as:
\begin{equation}
	\Delta = p_t-p_r. \label{eq32} 
\end{equation}

\section{Boundary condition} \label{sec6}
In the modified gravity theory, the boundary conditions applicable to a stellar system demand necessary modifications \cite{Rosa1}. The internal and external space-time regions of a $4-$ dimensional space-time manifold ($\Omega$) are separated across a $3-$ dimensional hypersurface, represented by $\Sigma$. It is determined by introducing the metric $h_{\alpha\beta}$ in the $X^l$ coordinate system. The projection tensors and normal vector are defined from $\Omega$ and $\Sigma$ in the form $e_\alpha ^a=\frac{\delta x^a}{\delta X^\alpha}$ and $n_a=\eta~\delta_a~l$, where the affine connection in the direction perpendicular to $\Sigma$ is denoted by $\ell$. The parameter $\eta$ takes the values -1,0, and 1 for time-like, null and space-like geodesics, respectively. Based on the structural configuration detailed above, the analysis proceeds by considering $n^ae_a ^\alpha=0$. Now, in the hypersurface ($\Sigma$), the form of induced metric $h_{\alpha\beta}$ and the extrinsic curvature tensor $K_{\alpha\beta}$ can be expressed as:
\begin{equation}
	h_{\alpha\beta}=e_\alpha ^a~ e_\beta ^b~ g_{ab}\\~~~~~~~~~~~~~~~~~~~~~~
	K_{\alpha\beta}=e_\alpha ^a~ e_\beta ^b~ \nabla_a n_b \label{eq33a} \\
\end{equation}
In the framework of $f(Q)$ gravity, the Schwarzschild spacetime is generally considered the most suitable candidate for representing the external spacetime surrounding compact objects. Given this connection, Wang \cite{wang} investigated the allowed forms of $f(Q)$ gravity in the context of static and spherically symmetric space-times with anisotropic matter. Their analysis revealed that the exact Schwarzschild solution is only possible when the second derivative of $f(Q)$ with respect to $Q$ is zero, and solutions obtained by considering a constant non-metricity scalar ($Q'=0$ or $Q=$constant) deviate from the solution of Schwarzschild. Consequently, to solve the field equations effectively for self-gravitating compact objects within the $f(Q)$ gravity framework, it's necessary to consider a functional form of $f(Q)$ where $f''(Q)$ itself is zero. Now, from the Eqs.~(\ref{eq21}), i.e. the off-diagonal component, one can derive:
\begin{eqnarray}
	f_{QQ}(Q)&=&0 , \nonumber \\
	f_Q(Q)&=&\alpha_0, \nonumber \\
	f(Q)&=&\alpha_0 +\alpha_1 Q, \label{eq34}
\end{eqnarray}
Now, in the exterior region, the corresponding energy-momentum tensor $T_{\mu \nu}=0$. So, the energy density $(\rho)$, radial pressure $(p_r)$, and tangential pressure $(p_t)$ does not exist in the exterior region. Therefore, substituting Eq.~(\ref{eq16}) in Eqs.~(\ref{eq18})-(\ref{eq20}) yields:
\begin{eqnarray}
	\nu'+\lambda'=0, \label{eq35} \\
	Q=\frac{\alpha_0}{\alpha_1}-\frac{2}{r^2}. \label{eq36}
\end{eqnarray}
Now, solving Eqs.~(\ref{eq16}), (\ref{eq35}) and (\ref{eq36}), we obtain,
\begin{equation}
	e^{-2\lambda}=1-\frac{c_1}{r}\bigg(\frac{1}{2\alpha_1}\bigg)-\bigg(\frac{\alpha_0}{6\alpha_1}\bigg)r^2 . \label{eq37}
\end{equation}
Then the line element can be written as:
\begin{equation}
	ds^2=-\bigg[1-\frac{1}{r}\bigg(\frac{c_1}{2\alpha_1}\bigg)-\frac{1}{3}\bigg(\frac{\alpha_0}{2\alpha_1}\bigg)r^2\bigg]dt^2+ \bigg[1-\frac{1}{r}\bigg(\frac{c_1}{2\alpha_1}\bigg)-\frac{1}{3}\bigg(\frac{\alpha_0}{2\alpha_1}\bigg)r^2 \bigg]^{-1}dr^2+ r^2(d\theta^2+\sin^2\theta d\phi^2) . \label{eq38}
\end{equation}
Further, refining Eq.~(\ref{eq38}) leads to,
\begin{equation}
	ds^2=-\bigg(1-\frac{2M}{r}-\frac{1}{3}\Lambda r^2\bigg)dt^2+\bigg(1-\frac{2M}{r}-\frac{1}{3}\Lambda r^2\bigg)^{-1}dr^2+ r^2\bigg(d\theta^2+\sin^2\theta d\phi^2\bigg) . \label{eq39}
\end{equation}
It resembles the Schwarzschild Anti-de Sitter exterior solution as described in GR, with $\Lambda=\frac{\alpha_0}{2\alpha_1}$ and $c_1=4M\alpha_1$, where $M$ denotes the mass of the star. Consequently, when we neglect the effect of cosmological constant $\Lambda$, it is evident that the Schwarzschild Anti-de Sitter space-time simplifies to the Schwarzschild exterior solution as given below:
\begin{equation}
	ds^2=-\bigg(1-\frac{2M}{r}\bigg)dt^2+\bigg(1-\frac{2M}{r}\bigg)^{-1}dr^2+r^2(d\theta^2 +\sin^2 \theta d\phi^2) . \label{eq33}
\end{equation}
To accurately model self-gravitating compact objects within the context of $f(Q)$ gravity with a linear functional form, it's crucial to select an appropriate exterior spacetime solution. This exterior solution should seamlessly match the interior solution at the boundary $(r=R)$ of the star where pressure vanishes. Given the nature of the linear $f(Q)$ gravity theory, the Schwarzschild spacetime emerges as the most suitable candidate for this exterior solution.\\
In the distribution formalism as developed by Rosa et al.\cite{Rosa}, the condition $[h_{ij} = 0]$ is imposed to ensure the continuity of the induced metric across the junction interface $(\Sigma)$. To uniquely identify the metric constant $(k)$, a matching condition must be satisfied. In the absence of a localised energy-momentum source, referred to as a "thin shell", the extrinsic curvature tensor must remain smooth across the boundary. This continuity is formally expressed as $[K_{ij} = 0]$ \cite{Rosa}. To derive the components of the extrinsic curvature for this configuration, we initiate the analysis using the Schwarzschild exterior line element, defined as Eq.~(\ref{eq33}).
In the parametric space of spherically symmetric space time, $k_{\alpha\beta}^{\pm}$ contains only two components which are $k_{tt}^{\pm}$ and $k_{rr}^{\pm}$  where, '$+$' and '$-$' stand for interior and exterior space-time respectively. Matching the Finch-Skea metric ansatz given in Eq.~(\ref{eq26}) with Eq.~(\ref{eq33}), we obtain the following components :
\begin{eqnarray}
	K_{tt}^+&=&\frac{M}{r^2\sqrt{1-\frac{2M}{r}}}, \nonumber \\
	K_{tt}^-&=& \frac{e^{\frac{2kr^2}{3}+\frac{(1+kr^2)^2(16B(n)\pi-\alpha_0)}{6k\alpha_1}r((1+kr^2)^2(16B(n)\pi-\alpha_0)+k(3+2kr^2)\alpha_1)}}{3(1+kr^2)^{7/6}\alpha_1}, \label{eqb1} \\
	K_{rr}^+ &=& -\frac{M}{r^2(1-\frac{2M}{r})^{5/2}}, \nonumber \\
	K_{rr}^- &=& \frac{kr}{\sqrt{1+kr^2}}, \label{eqb2}
\end{eqnarray}

A necessary condition for stellar equilibrium is that the radial pressure ($p_r$) should be decreasing monotonically with radius $(r)$ and must vanish at the stellar surface $(r = R)$, i.e,
\begin{equation}
	p_r(r=R)=0. \label{eq40}
\end{equation}
Now, the extrinsic curvature tensors are expressed in Eqs.~(\ref{eqb1})-(\ref{eqb2}) and using Eq.~(\ref{eq40}), we obtain,
\begin{equation}
	k=\frac{2M}{R^2(R-2M)}. \label{eq41}
\end{equation}
Once $k$ is known, the values of $\rho$, $p_r$ and $p_t$ inside the star can be evaluated.

\section{Mass-Radius Constraints for Compact Stars} \label{sec7}
The exact nature of the equation of state (EoS) of matter within a neutron star remains an unresolved enigma in astrophysics to date. This stems from the extreme densities within these stellar objects, often exceeding several times the nuclear saturation density, which are unattainable in a terrestrial laboratory. While the precise EoS of the interior matter of a neutron star remains elusive, the mass and radius of neutron stars estimated from astrophysical observations offer a valuable avenue for constraining its possible functional forms, or at least ruling out certain theoretical models. In this context, the mass-radius diagram derived from a specific EoS can be subjected to empirical validation through comparison with astrophysical observations.
\begin{figure}[h!]
	\centering
	\includegraphics[width=10cm]{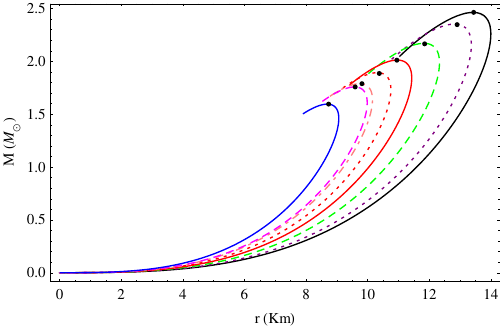}
	\caption{ Mass- radius plot for different values of $B$. Lines from top to bottom represent $n=1.110,~0.871,~0.740,~0.660,~0.608,~0.563,~0.552$ and $0.478~fm^{-3}$. Maximum mass points are indicated through solid dots.}
	\label{figc}
\end{figure}
\begin{table*}[t!]
	\centering
	\caption{Maximum mass-radius prediction from the TOV equations.}
	\label{tab:1}
	\begin{tabular}{|cccc|}
		\hline
		$n(fm^{-3})$ &  $B(n) (MeV/fm^3)$ & $M_{max}(M_{\odot})$ & $R_{max}(Km)$ \\ \hline
		$0.478$ & $91.55$ & $1.59$ & $9.06$ \\
		$0.552$ & $75.42$ & $1.75$ & $9.98$ \\ 
		$0.563$ & $72.90$ & $1.78$ & $10.15$ \\ 
		$0.608$ & $65.02$ & $1.89$ & $10.75$ \\ 
		$0.660$ & $57.55$ & $2.01$ & $11.43$ \\
		$0.740$ & $49.45$ & $2.17$ & $12.33$ \\
		$0.871$ & $42.10$ & $2.35$ & $12.82$ \\ 
		$1.40$ & $38.01$ & $2.47$ & $14.07$ \\ 
		\hline
	\end{tabular}
\end{table*}
We have evaluated, by solving TOV equations, the maximum mass and radius, and these are shown in Fig.~\ref{figc}. Further, the results are also presented in a tabular form in Table~\ref{tab:1}. It is noted that for different values of baryon number density $(n)$, the corresponding maximum mass varies significantly. If one increases the baryon number density $(n)$ from $0.478~fm^{-3}$ to $1.40~fm^{-3}$, it is observed that the maximum mass also increases in this model. From Fig.~\ref{figb} and Table~\ref{tab:1}, it is noted that for large value of $n~(=1.40~fm^{-3})$, $B(n)$ approaches to a constant value of $38.01~MeV/fm^3$. It is also noted that the model can accommodate a maximum mass of $2.47~M_\odot$ for $B(n)=38.01~MeV/fm^3$. Now, according to Bodmer and Witten \cite{Witten} hypothesis, the maximum mass ranging from $1.59-2.01~M_\odot$ associated with the range of $B(n)$ as $91.55-57.55~MeV/fm^3$ may be treated as SS. On the other hand, maximum mass above $2.01~M_\odot$ and up to $2.47~M_\odot$ corresponding to the value of $B(n)$ less than $57.55~MeV/fm^3$ and up to $38.01~MeV/fm^3$ may be treated as diquark star as Bodmer and Witten \cite{Witten} hypothesis states that below the value of $B(n)=57.55~MeV/fm^3$, quark matter composed only of two types of flavours up $(u)$ and down $(d)$. Such a diquark star, however, may be stable in the presence of strong interaction. 

\section{Physical analysis of the proposed model} \label{sec8}
In the proposed model, we consider a well-behaved and non-singular metric potential, denoted by $\lambda(r)$ as proposed by Finch-Skea \cite{Finch}. Solution of Eq.~(\ref{eq24}), yields the another metric potential $\nu(r)$. Notably, $\nu(r)$ remains finite at the centre. Both the metric potentials exhibit regular as well as non-singular behaviour throughout the stellar configuration. Consequently, the obtained solution for the stellar system is physically acceptable. In order to check the viability of the proposed model, we analyse the graphical nature with respect to radial distance.  We have considered $4U1820-30$ with mass of $1.58~M_{\odot}$ and a radius of $9.1~Km$ \cite{Tolga} to discuss the physical acceptability of that model. For this purpose, we analyse the radial variations of $\rho$, $p_r$, $p_t$, $\Delta$ and are shown graphically. We have also studied the causality conditions and energy conditions. Apart from the above analysis, it is interesting to note that the estimated radius of $4U~1820-30$ from observation may be precisely predicted from the present model, considering the non-metricity parameter $\alpha_0$ and $\alpha_1$, and the results are tabulated in Table~\ref{tab:4}. 
\begin{table}[h!]
	\centering
	\caption{Predicted radius of $4U1820-30$ for different baryon number density $(n)$ and $\alpha_0=1\times10^{-3}~Km^{-2}$, $\alpha_1=~-0.75$.}
	\label{tab:4}
	\begin{tabular}{|ccc|}
		\hline
		Baryon number density & Bag parameter & Predicted radius  \\ 
		$(n~fm^{-3})$ & $B(n)~(MeV/fm^3)$ & $(Km)$ \\	\hline
		$0.660$ & $57.55$ & $10.05$  \\
		$0.563$ & $72.90$ & $9.10$  \\
		$0.478$ & $91.55$ & $8.25$  \\ \hline
		
	\end{tabular}
\end{table}
\begin{table}[h!]
	\centering
	\caption{Radius prediction of $4U~1820-30$ for different choices of $\alpha_0$ and $n=0.5~fm^{-3}$, $\alpha_1=-1.15$.}
	\label{tab:5}
	\begin{tabular}{|cc|}
		\hline
		$\alpha_0$ & Predicted radius  \\ 
		$(Km^{-2}) $ & $ (Km) $ \\ \hline
		$1\times10^{-3}$ & $9.10$ \\
		$3\times10^{-3}$ & $9.43$ \\
		$5\times10^{-3}$ & $9.80$ \\ \hline
	\end{tabular}
\end{table}
It is also studied that for a fixed value of $n$ and $\alpha_1$, how the radius of a star predicted from the model, changes with $\alpha_0$ and is shown in Table~\ref{tab:5}. We obtain the basic physical parameters such as central density $(\rho_c)$, surface density $(\rho_s)$ and central pressure $(p_c)$ of some compact stars and are shown in Table~\ref{tab:3}.
\begin{table*}[t!]
	\centering
	\caption{Tabulation of some physical parameters of some known compact objects for $n=0.578~fm^{-3}$}
	\label{tab:3}
	\begin{tabular}{|cccc|}
		\hline
		Compact objects & Central density ($\rho_c$) & Surface density($\rho_s$) & Central pressure($p_c$) \\
		& $(g/cc)$ & $(g/cc)$ & $(dyne/cm^3)$ \\ \hline
		4U 1820-30 \cite{Tolga} & $0.893\times10^{15}$ & $0.342\times10^{15}$ & $1.19\times10^{35}$ \\
		Her X-1 \cite{Abubekerov} & $0.565\times10^{15}$ & $0.330\times10^{15}$ & $0.20.\times10^{35}$ \\
		LMC X-4 \cite{Rawls} & $0.624\times10^{15}$ & $0.330\times10^{15}$ & $0.38\times10^{35}$ \\ 
		SMC X-4 \cite{Rawls} & $0.720\times10^{15}$ & $0.332\times10^{15}$ & $0.67\times10^{35}$ \\
		Cen X-3 \cite{Rawls} & $0.828\times10^{15}$ & $0.338\times10^{15}$ & $0.99\times10^{35}$  \\
		Vela X-1 \cite{Rawls} & $1.128\times10^{15}$ & $0.367\times10^{15}$ & $1.89\times10^{35}$ \\
		EXO 1745-248 \cite{Ozel} & $0.576\times10^{15}$ & $0.269\times10^{15}$ & $0.235\times10^{35}$ \\
		4U 1608-52 \cite{Tolga1} & $1.070\times10^{15}$ & $0.360\times10^{15}$ & $1.72\times10^{35}$ \\ 
		\hline
	\end{tabular}
\end{table*}

\subsection{Energy density and pressure profiles} 
\begin{figure}[h!]
	\centering
	\begin{subfigure}{0.48\textwidth}
		\centering
		\includegraphics[scale=0.8]{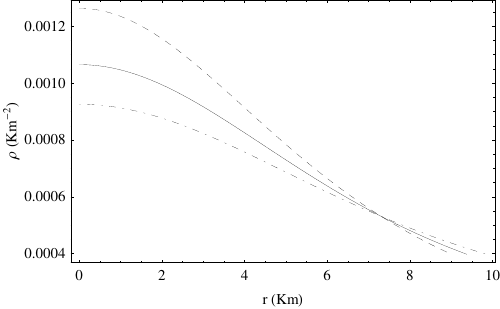}
		\caption{Lines from top to bottom represent $\alpha_0 = 5\times10^{-3},~ 3\times10^{-3}$ and $1\times10^{-3}~Km^{-2}$ respectively.}
		\label{fig1a}
	\end{subfigure}
	\hfill
	\begin{subfigure}{0.45\textwidth}
		\centering
		\includegraphics[scale=0.8]{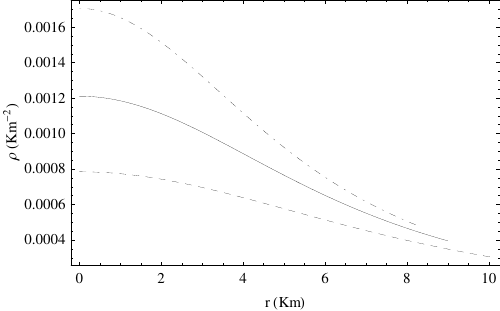}
		\caption{Lines from top to bottom represent $n=0.478,~0.563$ and $0.660~fm^{-3}$ respectively.}
		\label{fig1b}
	\end{subfigure}
	\caption{Variation of energy density $(\rho)$ with radius $(r)$.}
	\label{fig1}
\end{figure}

\begin{figure}[h!]
	\centering
	\begin{subfigure}[b]{0.45\textwidth}
		\centering
		\includegraphics[width=\textwidth]{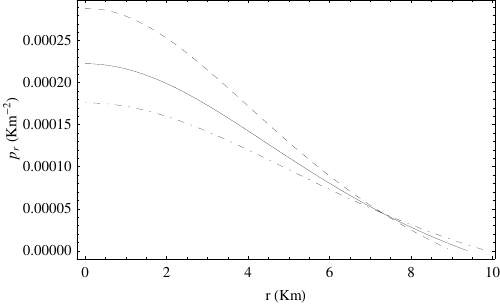}
		\caption{Lines from top to bottom represent $\alpha_0 = 5\times10^{-3},~ 3\times10^{-3}$ and $1\times10^{-3}~Km^{-2}$ respectively.}
		\label{fig2a}
	\end{subfigure}
	\hfill
	\begin{subfigure}[b]{0.45\textwidth}
		\centering
		\includegraphics[width=\textwidth]{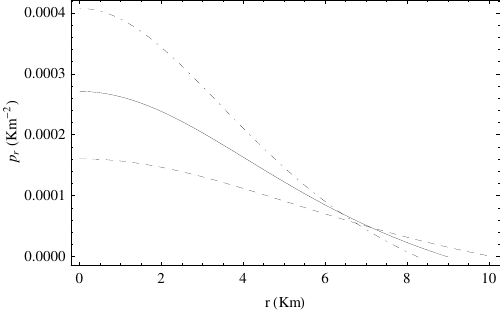}
		\caption{Lines from top to bottom represent $n=0.478,~0.563$ and $0.660~fm^{-3}$ respectively.}
		\label{fig2b}
	\end{subfigure}
	\caption{Variation of radial pressure $(p_{r})$ with radius $(r)$.}
	\label{fig2}
\end{figure}

\begin{figure}[h!]
	\centering
	\begin{subfigure}{0.48\textwidth}
		\centering
		\includegraphics[scale=0.8]{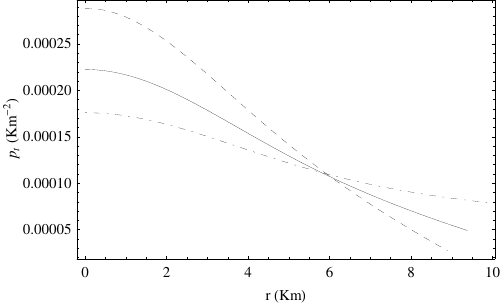}
		\caption{Lines from top to bottom represent $\alpha_0 = 5\times10^{-3},~ 3\times10^{-3}$ and $1\times10^{-3}~Km^{-2}$ respectively.}
		\label{fg3a}
	\end{subfigure}
	\hfill
	\begin{subfigure}{0.48\textwidth}
		\centering
		\includegraphics[scale=0.8]{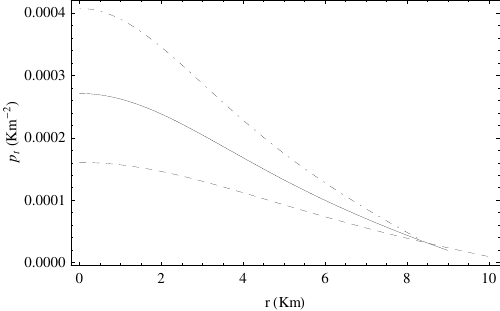}
		\caption{Lines from top to bottom represent $n=0.478,~0.563$ and $0.660~fm^{-3}$ respectively.}
		\label{fig3b}
	\end{subfigure}
	\caption{Variation of tangential pressure $(p_{t})$ with radius $(r)$.}
	\label{fig3}
\end{figure}

In this section, we analyse the physical behaviour of the model by studying the graphical nature of the energy density profile, radial as well as tangential pressure profiles. To sustain a physically stable structure, the energy density and pressure must have a monotonically decreasing nature with respect to radius. Fig.~\ref{fig1} illustrates the radial variation of the energy density ($\rho$), which represents a monotonically decreasing nature. Fig.~\ref{fig1a} represents that the $f(Q)$-gravity parameter $\alpha_0$ has a significant effect on energy density profile. It is observed that the energy density increases as we gradually increase $\alpha_0$ from $1\times10^{-3}$ to  $5\times10^{-3}~Km^{-2}$. Fig.~\ref{fig1b} shows how the energy density changes with the baryon number density ($n$). The energy density ($\rho$) falls gradually when the baryon number density $(n)$ increases from $0.478$ to $0.660~fm^{-3}$. The radial pressure $(p_r)$ is shown in Fig.~\ref{fig2}. As one goes from the centre of the star to the surface, the radial pressure ($p_r$) falls off and finally vanishes at the stellar surface. Fig.~\ref{fig2a} indicates that the radial pressure $(p_r)$ increases when parameter $\alpha_0$ increases. The effect caused by the baryon number density $(n)$ demonstrates that the radial pressure decreases as the baryon number density $(n)$ increases, as shown in Fig.~\ref{fig2b}. Same findings hold for the tangential pressure $(p_t)$ as shown in Fig.~\ref{fig3}. \\

\subsection{Behaviour of pressure anisotropy}
\begin{figure}[ht!]
	\centering
	\begin{subfigure}{0.48\textwidth}
		\centering
		\includegraphics[scale=0.8]{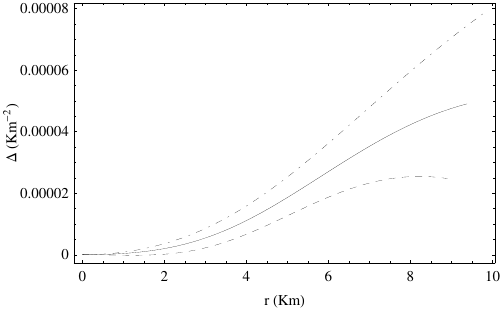}
		\caption{Lines from top to bottom represent $\alpha_0 = 5\times10^{-3},~ 3\times10^{-3}$ and $1\times10^{-3}~Km^{-2}$ respectively.}
		\label{fig4a}
	\end{subfigure}
	\hfill
	\begin{subfigure}{0.48\textwidth}
		\centering
		\includegraphics[scale=0.8]{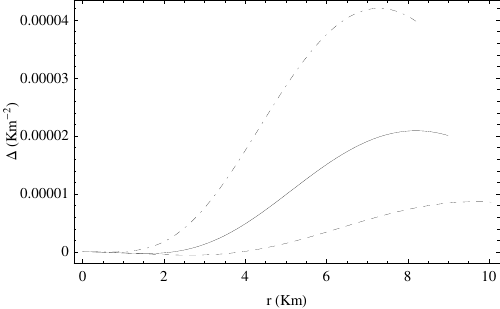}
		\caption{Lines from top to bottom represent $n=0.478,~0.563$ and $0.660~fm^{-3}$ respectively.}
		\label{fig4b}
	\end{subfigure}
	\caption{Variation of anisotropy $(\Delta)$ with radius $(r)$.}
	\label{fig4}
\end{figure}

A repulsive force comes from anisotropy when the tangential pressure dominates the radial pressure \cite{Gokhroo}. Such repulsive force originated from pressure anisotropy, counterbalances the inward pull due to the gravitational attractive force, thereby increasing the overall stability of the system.  From Fig.~\ref{fig4}, we note that anisotropy is increasing with the increase of radial distance from the centre. We also observed from Fig.~\ref{fig4a} that at the stellar surface, anisotropy increases with the increase of $\alpha_0$. Additionally, we emphasise that the anisotropy can also be governed by the baryon number density ($n$). An increase in $n$ is accompanied by a decrease in the anisotropy parameter ($\Delta$), as shown in Fig.~\ref{fig4b}.

\subsection{Causality condition}
\begin{figure}[h!]
	\centering
	\begin{subfigure}{0.48\textwidth}
		\centering
		\includegraphics[scale=0.8]{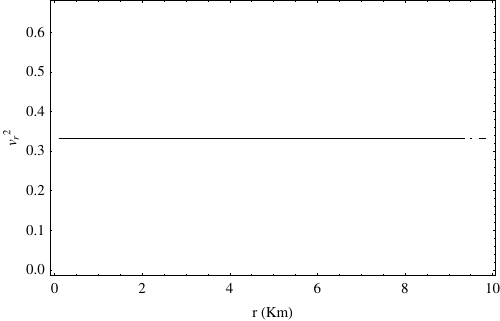}
		\caption{Lines from top to bottom represent $\alpha_0 = 5\times10^{-3},~ 3\times10^{-3}$ and $1\times10^{-3}~Km^{-2}$ respectively.}
		\label{fig5a}
	\end{subfigure}
	\hfill
	\begin{subfigure}{0.48\textwidth}
		\centering
		\includegraphics[scale=0.8]{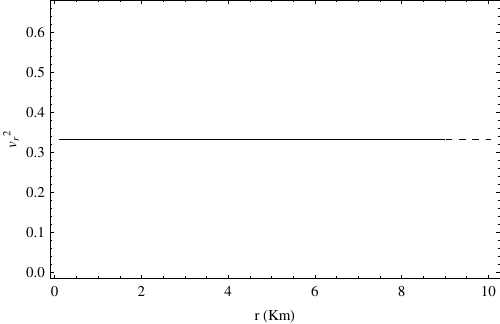}
		\caption{Lines from top to bottom represent $n=0.478,~0.563$ and $0.660~fm^{-3}$ respectively.}
		\label{fig5b}
	\end{subfigure}
	\caption{Variation of sound parameter $v_{r}^2$ with radius $(r)$.}
	\label{fig5}
\end{figure}

\begin{figure}[h!]
	\centering
	
	\begin{subfigure}{0.48\textwidth}
		\centering
		\includegraphics[scale=0.8]{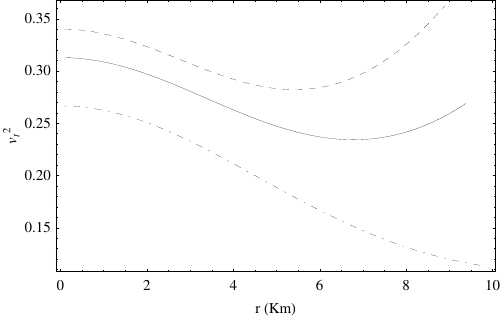}
		\caption{Lines from top to bottom represent $\alpha_0 = 5\times10^{-3},~ 3\times10^{-3}$ and $1\times10^{-3}~Km^{-2}$ respectively.}
		\label{fig6a}
	\end{subfigure}
	\hfill
	\begin{subfigure}{0.48\textwidth}
		\centering
		\includegraphics[scale=0.8]{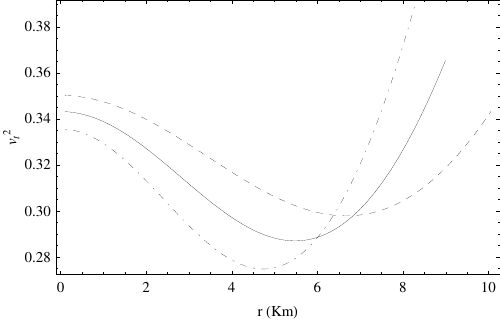}
		\caption{Lines from top to bottom represent $n=0.478,~0.563$ and $0.660~fm^{-3}$ respectively.}
		\label{fig6b}
	\end{subfigure}
	
	\caption{Variation of sound parameter $v_{t}^2$ with radius $(r)$.}
	\label{fig6}
\end{figure}

For a physically viable model of an anisotropic fluid sphere, the radial and tangential sound speeds must not exceed the speed of light. This requirement is commonly referred to as the \textit{causality condition}. The concepts of causality and stability in a stellar configuration are closely connected, since a system becomes unstable if any propagation mode travels faster than light (here we take $c=1$). Therefore, verifying the causality condition is essential when analysing a stellar structure to ensure its physical stability. The radial $(v_r^2)$ and transverse $(v_t^2)$ sound speeds are given by:
\begin{eqnarray}
	v_r^2&=&\frac{dp_r}{d\rho} , \nonumber \\
	v_t^2&=&\frac{dp_t}{d\rho} . \label{eq42}
\end{eqnarray}
The causality requirement places a strict upper limit on the sound speeds such that $v_r^2 \leq 1$ and $v_t^2 \leq 1$ under the relativistic unit system where $\hbar = 1$ and $c = 1$. Moreover, thermodynamic stability demands that both sound speeds remain positive, i.e., $v_r^2 > 0$ and $v_t^2 > 0$. For a stellar model to be physically realistic, these conditions must hold simultaneously throughout the interior region of the star. Therefore, the combined constraints can be written as:
\begin{equation}
	0<v_r^2\leq 1 \nonumber \\
\end{equation}
and
\begin{equation}
	0<v_t^2\leq1 . \nonumber
\end{equation}
Because the analytical expressions for the sound speeds are quite complicated, we illustrate their radial behaviour graphically in Figs.~\ref{fig5} and \ref{fig6}.

\subsection{Energy condition}
\begin{figure}[h!]
	\centering
	
	\begin{subfigure}{0.48\textwidth}
		\centering
		\includegraphics[scale=0.8]{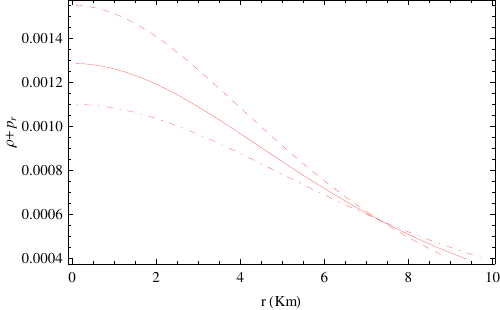}
		\caption{Lines from top to bottom represent $\alpha_0 = 5\times10^{-3},~ 3\times10^{-3}$ and $1\times10^{-3}~Km^{-2}$ respectively.}
		\label{fig7a1}
	\end{subfigure}
	\hfill
	\begin{subfigure}{0.48\textwidth}
		\centering
		\includegraphics[scale=0.8]{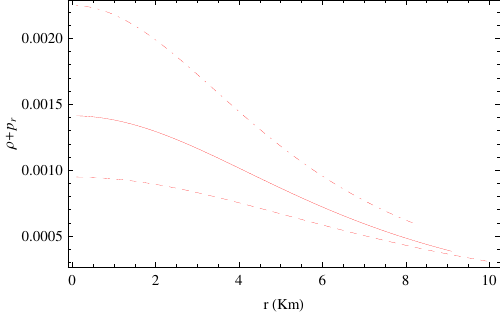}
		\caption{Lines from top to bottom represent $n=0.478,~0.563$ and $0.660~fm^{-3}$ respectively.}
		\label{fig7b1}
	\end{subfigure}
	
	\caption{Variation of $(\rho + p_r)$ with radius $(r)$.}
	\label{fig7a}
\end{figure}

\begin{figure}[h!]
	\centering
	
	\begin{subfigure}{0.48\textwidth}
		\centering
		\includegraphics[scale=0.8]{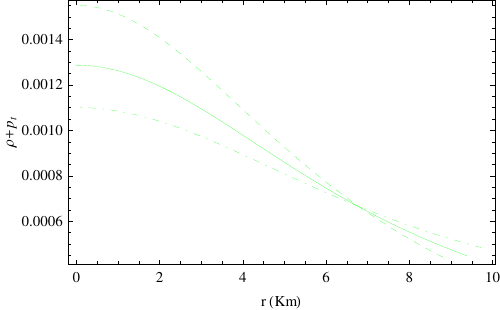}
		\caption{Lines from top to bottom represent $\alpha_0 = 5\times10^{-3},~ 3\times10^{-3}$ and $1\times 10^{-3}~Km^{-2}$ respectively.}
		\label{fig7a2}
	\end{subfigure}
	\hfill
	\begin{subfigure}{0.48\textwidth}
		\centering
		\includegraphics[scale=0.8]{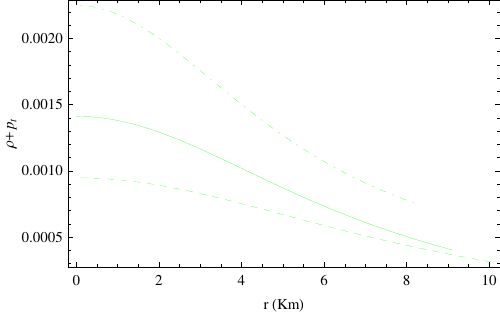}
		\caption{Lines from top to bottom represent $n=0.478,~0.563$ and $0.660~fm^{-3}$ respectively.}
		\label{fig7b2}
	\end{subfigure}
	
	\caption{Variation of $(\rho + p_t)$ with radius $(r)$.}
	\label{fig7b}
\end{figure}

\begin{figure}[h!]
	\centering
	
	\begin{subfigure}{0.48\textwidth}
		\centering
		\includegraphics[scale=0.8]{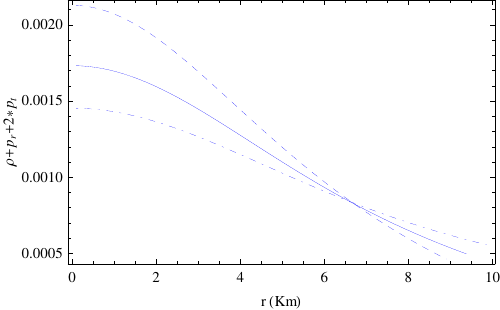}
		\caption{Lines from top to bottom represent $\alpha_0 = 5\times10^{-3},~ 3\times10^{-3}$ and $1\times10^{-3}~Km^{-2}$ respectively.}
		\label{fig7a3}
	\end{subfigure}
	\hfill
	\begin{subfigure}{0.48\textwidth}
		\centering
		\includegraphics[scale=0.8]{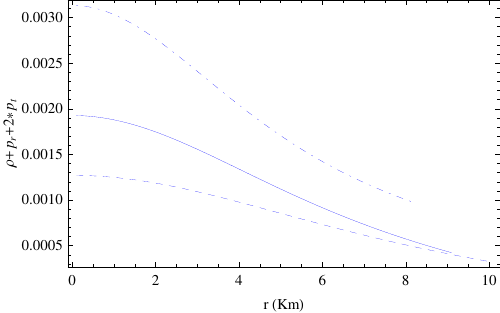}
		\caption{Lines from top to bottom represent $n=0.478,~0.563$ and $0.660~fm^{-3}$ respectively.}
		\label{fig7b3}
	\end{subfigure}
	
	\caption{Variation of $(\rho + p_r + 2p_t)$ with radius $(r)$.}
	\label{fig7c}
\end{figure}

\begin{figure}[h!]
	\centering
	
	\begin{subfigure}{0.48\textwidth}
		\centering
		\includegraphics[scale=0.8]{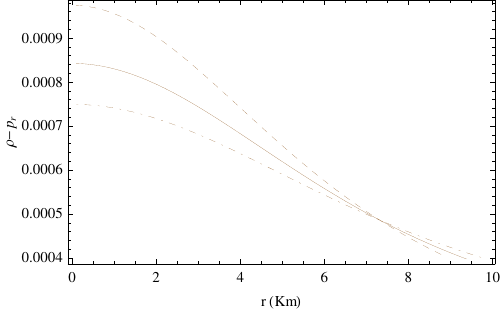}
		\caption{Lines from top to bottom represent $\alpha_0 = 5\times10^{-3},~ 3\times10^{-3}$ and $1\times10^{-3}~Km^{-2}$ respectively.}
		\label{fig7a4}
	\end{subfigure}
	\hfill
	\begin{subfigure}{0.48\textwidth}
		\centering
		\includegraphics[scale=0.8]{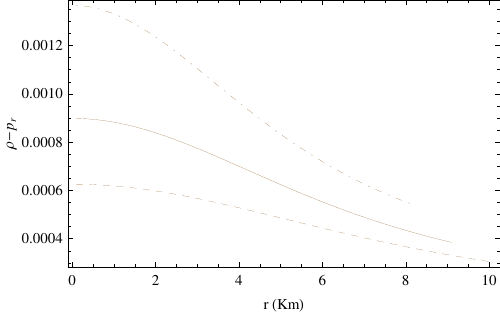}
		\caption{Lines from top to bottom represent $n=0.478,~0.563$ and $0.660~fm^{-3}$ respectively.}
		\label{fig7b4}
	\end{subfigure}
	
	\caption{Variation of $(\rho - p_r)$ with radius $(r)$.}
	\label{fig7d}
\end{figure}

\begin{figure}[h!]
	\centering
	
	\begin{subfigure}{0.48\textwidth}
		\centering
		\includegraphics[scale=0.8]{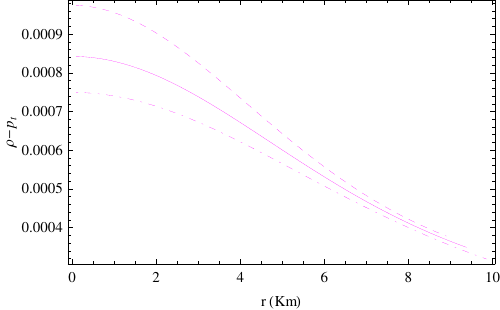}
		\caption{Lines from top to bottom represent $\alpha_0 = 5\times10^{-3},~ 3\times10^{-3}$ and $1\times10^{-3}~Km^{-2}$ respectively.}
		\label{fig7a5}
	\end{subfigure}
	\hfill
	\begin{subfigure}{0.48\textwidth}
		\centering
		\includegraphics[scale=0.8]{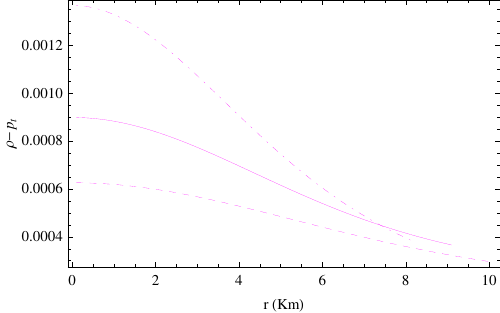}
		\caption{Lines from top to bottom represent $n=0.478,~0.563$ and $0.660~fm^{-3}$ respectively.}
		\label{fig7b5}
	\end{subfigure}
	
	\caption{Variation of $(\rho - p_t)$ with radius $(r)$.}
	\label{fig7e}
\end{figure}

One of the most important properties that should be satisfied by any fluid configuration is the fulfilment of energy conditions \cite{Kolassis,Hawking,Wald}. These conditions will be satisfied if the following inequalities hold simultaneously:
\begin{enumerate}
	\item Null energy condition (NEC): $\rho+p_{r}\geq 0,~\rho+p_{t}\geq 0$\\
	\item Weak energy condition (WEC): $\rho\geq 0,~\rho+p_{r}\geq 0,~\rho+p_{t}\geq 0$.\\
	\item Strong energy condition (SEC): $\rho+p_{r}\geq 0,~\rho+p_{t}\geq0,~ \rho+p_{r}+2p_{t}\geq0$.\\
	\item Dominent energy condition (DEC): $\rho\geq0,~\rho-p_{r}\geq0,~\rho-p_{t}\geq0$\\
\end{enumerate}
We analyse the energy conditions in the present context with the help of graphical representations. From Fig.~\ref{fig1} and Figs.~\ref{fig7a} - \ref{fig7e}, it is observed that the proposed model satisfies the NEC, WEC, SEC and DEC.
The effect of parameter $\alpha_0$ and the baryon number density $(n)$ on the energy conditions has also been satisfied, and from Fig.~\ref{fig1} and Figs.~\ref{fig7a} -\ref{fig7e}, we note that $\alpha_0$ and $n$ have some distinct effects on energy conditions.
\section{Stability analysis} \label{sec9}
The stability of any stellar structure ensures that the equilibrium of the proposed model is physically viable and will not be disturbed by any external disturbance. The stability of our proposed model is analysed based on the following methods:
\begin{enumerate}
	\item Generalised TOV equation.
	\item Herrera's cracking condition.
	\item Study of adiabatic index of fluid sphere.
\end{enumerate}

\subsection{Generalised TOV equation}
Equilibrium state of a fluid sphere under three forces, i.e. gravitational, hydrostatic and anisotropic forces, can be analysed to determine whether they satisfy the generalised Tolman-Oppenheimer-Volkoff (TOV) \cite{Tolman,Oppenheimer} equation or not. The generalised TOV equation is given by:
\begin{equation}
	-\frac{M_{G}(r)(\rho+p_{r})}{r^2}e^{(\lambda -\nu)}-\frac{dp_{r}}{dr}+\frac{2\Delta}{r}=0 , \label{eq43}
\end{equation} 
where $M_{G}$ is represented as the active gravitational mass derived from the mass formula of Tolman-Whittaker \cite{Gron} given as:
\begin{equation}
	M_{G}(r)=r^2\nu'e^{(\nu-\lambda)}. \label{eq44}
\end{equation}
Using the value of $M_{G}$ in Eq.~(\ref{eq43}), we obtain:
\begin{equation}
	-\nu'(\rho+p_{r})-\frac{dp_r}{dr}+\frac{2\Delta}{r}=0 . \label{eq45}
\end{equation} 
The above expression may also be written as:
\begin{equation}
	F_g + F_h +F_a =0 , \label{eq46}
\end{equation}
where the terms $F_g,~F_h$ and $F_a$ represent, respectively, the gravitational, hydrostatic and anisotropic forces and can be expressed as follows:
\begin{eqnarray}
	F_g&=&-\nu'(\rho +p_{r}), \label{eq47} \\
	F_{h}&=&-\frac{dp_r}{dr}, \label{eq48}\\
	F_{a}&=&\frac{2\Delta}{r}. \label{eq49}
\end{eqnarray}
\begin{figure}[ht!]
	\centering
	\begin{subfigure}{0.48\textwidth}
		\centering
		\includegraphics[scale=0.8]{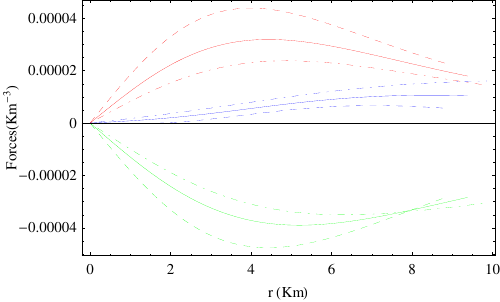}
		\caption{Lines from top to bottom represent $\alpha_0 = 5\times10^{-3},~ 3\times10^{-3}$ and $1\times10^{-3}~Km^{-2}$ respectively.}
		\label{fig8a}
	\end{subfigure}
	\hfill
	\begin{subfigure}{0.48\textwidth}
		\centering
		\includegraphics[scale=0.8]{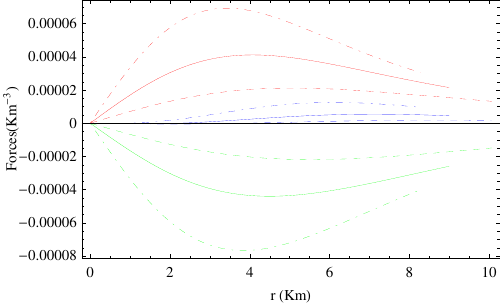}
		\caption{Lines from top to bottom represent $n=0.478,~0.563$ and $0.660~fm^{-3}$ respectively.}
		\label{fig8b}
	\end{subfigure}
	
	\caption{TOV equation profile of 4U1820-30.}
	\label{fig8}
\end{figure}
Here, all the forces are in units of $Km^{-3}$. It is observed that although these forces individually contribute to either an attractive or a repulsive nature, the combined effect of these three different forces shows that the sum of the three forces described above is zero inside as well as on the surface of the star, satisfying the Tolman-Oppenheimer-Volkoff (TOV) equation. The profiles of the three different forces are plotted in Fig.~\ref{fig8} in hydrostatic equilibrium.

\subsection{Cracking condition proposed by Herrera}
Any stellar model should also be stable with respect to small fluctuations in their physical variables. To check the stability of a stellar model, the concept of 'cracking' was introduced by Herrera \cite{Herrera}. Based on Herrera's concept, Abreu et al. \cite{Abreu} introduced a criterion which states that a stellar model will be in stable equilibrium if the square of radial velocity $(v_r^2)$ and tangential velocity $(v_t^2)$ obey the following condition,
\begin{equation}
	0\leq |v_r^2-v_t^2|\leq 1 . \label{eq50}
\end{equation}
A stellar model stands by with the relation Eq.~(\ref{eq50}), which is considered a stable one.
\begin{figure}[ht!]
	\centering
	
	\begin{subfigure}{0.48\textwidth}
		\centering
		\includegraphics[scale=0.8]{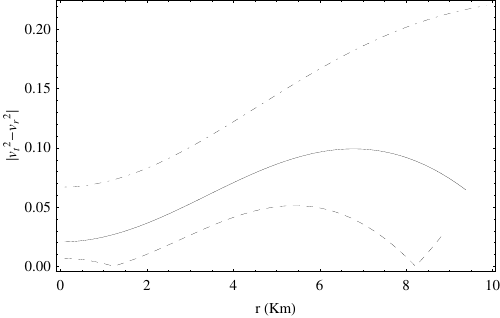}
		\caption{Lines from top to bottom represent $\alpha_0 = 5\times10^{-3},~ 3\times10^{-3}$ and $1\times10^{-3}~Km^{-2}$ respectively.}
		\label{fig9a}
	\end{subfigure}
	\hfill
	\begin{subfigure}{0.48\textwidth}
		\centering
		\includegraphics[scale=0.8]{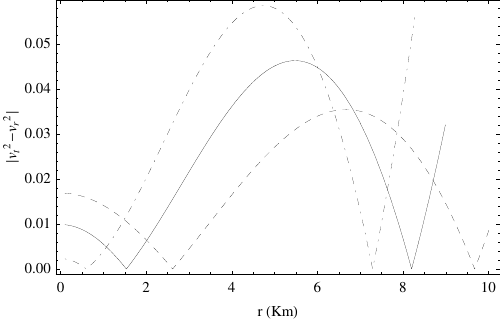}
		\caption{Lines from top to bottom represent $n=0.478,~0.563$ and $0.660~fm^{-3}$ respectively.}
		\label{fig9b}
	\end{subfigure}
	
	\caption{Variation of $|v_r^2 - v_t^2|$ with radius $(r)$.}
	\label{fig9}
\end{figure}
From the graphical nature as shown in the Fig.~\ref{fig9}, it is noted that the Abreu inequality given in Eq.~(\ref{eq50}) is satisfied at all interior points. 

\subsection{Radial variation of Adiabatic index}
\begin{figure}[h!]
	\centering
	
	\begin{subfigure}{0.48\textwidth}
		\centering
		\includegraphics[scale=0.8]{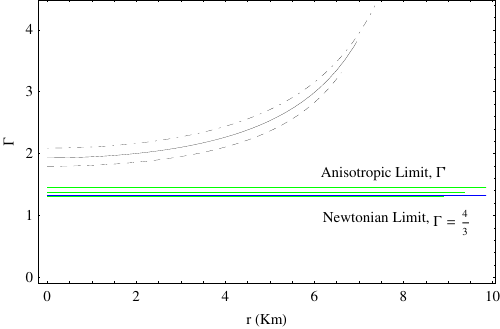}
		\caption{Lines from top to bottom represent $\alpha_0 = 5\times10^{-3},~ 3\times10^{-3}$ and $1\times10^{-3}~Km^{-2}$ respectively.}
		\label{fig10a}
	\end{subfigure}
	\hfill
	\begin{subfigure}{0.48\textwidth}
		\centering
		\includegraphics[scale=0.8]{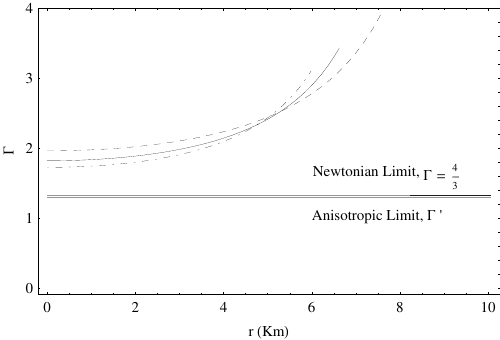}
		\caption{Lines from top to bottom represent $n=0.478,~0.563$ and $0.660~fm^{-3}$ respectively.}
		\label{fig10b}
	\end{subfigure}
	
	\caption{Variation of Adiabatic index $(\Gamma)$ with radius $(r)$.}
	\label{fig10}
\end{figure}

The stability of a relativistic fluid sphere with anisotropic pressures is linked to its adiabatic index ($\Gamma$). The ratio of two specific heats, expressed as \cite{Chan}:
\begin{equation}
	\Gamma=\frac{\rho+p_{r}}{p_{r}}\bigg(\frac{dp_{r}}{d\rho}\bigg)=\bigg(\frac{\rho+p_{r}}{p_{r}}\bigg)v_{r}^2 . \label{eq51}
\end{equation} 
For Newtonian fluid, the equilibrium is characterised by $\Gamma =\frac{4}{3}$, as proposed by Heintzmann \cite{Heintzman}. In a relativistic isotropic sphere, this criterion is modified because the pressure’s restorative effect tends to increase instability. In the case of a general relativistic sphere composed of anisotropic fluid, the stability analysis is more complex, as it depends on the specific nature of the anisotropy. Such a condition is modified by Chan et al. \cite{Chan1} and shows that within the parameter space used here, the condition of stability can be written as: 
\begin{equation}
	\Gamma\geq\Gamma' , \label{eq52}
\end{equation}
where,
\begin{equation}
	\Gamma' =\frac{4}{3}-\bigg[\frac{4}{3}\frac{(p_{r}-p_{t})}{|p_{r}'|r}\bigg]_{max} . \label{eq53}
\end{equation}
For the present model, the profile of adiabatic index ($\Gamma$) is shown in Fig.~\ref{fig10} for different values of $n$. From Fig.~\ref{fig10}, it is noted that the condition imposed by Chan et al. \cite{Chan1} is well satisfied throughout the interior of the stellar configuration and hence the condition of stability is satisfied. 
\newpage
\section{Prediction of radii of compact stars} \label{sec10}
In the present model, the radii of several known compact stars are estimated and subsequently compared with the radii obtained from the TOV approach. The corresponding results are presented in Table~\ref{tab:2}.
\begin{table*}[t!]
	\centering
	\caption{predicted radius of some compact object from TOV equations.}
	\label{tab:2}
	\begin{tabular}{|ccc|cc|cc|cc|}
		\hline
		Compact  &  Measured  & Measured  & \multicolumn{6}{c|}{ Predicted radius from} \\ \cline{4-9}
		object & mass  & radius  & \multicolumn{2}{c|}{TOV}  &  \multicolumn{4}{c|}{Model}   \\ \cline{6-9}
		& $(M/M_{\odot})$ & $(Km)$ & & & \multicolumn{2}{c|}{$\alpha_0=~1\times10^{-3}~(Km^{-2})$} & \multicolumn{2}{c|}{$B(n)=75~MeV/fm^3$} \\ \cline{4-9}
		& &  & $B(n)$ & $R$ & $B(n)$ & $R$ & $\alpha_0$ & $R$ \\ 
		& & & $(MeV/fm^3)$ & $(Km)$ & $(MeV/fm^3)$ & $(Km)$ & $(Km^{-2})$ & $(Km)$ \\ \hline
		Her X-1 \cite{Abubekerov} & $0.85\pm 0.15$ & $8.1\pm 0.41$ & $91.55$ & $8.11$ & $67.00$ & $8.11$ & $3.1\times10^{-3}$ & $8.11$ \\ 
		LMC X-4 \cite{Rawls} & $1.04\pm0.09$ & $8.301\pm 0.2$  & $91.55$ & $8.55$ & $72.00$ & $8.32$ & $2\times10^{-3}$ & $8.35$ \\
		SMC X-4 \cite{Rawls} & $1.29\pm0.05$ & $8.823\pm 0.09$ & $91.55$ & $8.95$  & $71.00$ & $8.81$ & $2\times10^{-3}$ & $8.80$ \\
		EXO 1745-248 \cite{Ozel} & $1.4 $ & $11.00$ &  $57.55$ & $10.96$ & $57.55$ & $9.78$ & $9.5\times10^{-3}$ & $11.04$ \\
		Cen X-3 \cite{Rawls} & $1.49\pm 0.08$ & $9.178\pm0.13$ & $90.00$ & $9.14$ & $70.00$ & $9.15$ & $2.5\times10^{-3}$ & $9.18$ \\
		4U 1820-30 \cite{Tolga} & $1.58\pm0.06$ & $9.1\pm0.4$ & $88.00$ & $9.17$ & $72$ & $9.15$ & $1.6\times10^{-3}$ & $9.11$ \\ 
		4U 1608-52 \cite{Tolga1} & $1.74\pm 0.14$ & $9.528\pm0.15$ & $77.00$ & $9.47$ & $68.00$ & $9.55$ & $2.8\times10^{-3}$ & $9.53$ \\ 
		Vela X-1 \cite{Rawls}& $1.77\pm 0.08$ & $9.56\pm 0.08$ & $74.00$ & $9.79$ & $68.00$ & $9.58$ & $2.8\times10^{-3} $ & $9.56$ \\   \hline
	\end{tabular}
\end{table*}
It is interesting to note that the estimated radii of such stars from recent observations may also be precisely predicted with a suitable parametric choice of model parameters $n,~B,~\alpha_0$ and $\alpha_1$. Thus, the $f(Q)$ modified theory of gravity with linear form may be useful to study the compact stars.
\section{Discussion} \label{sec11}
We investigate the characteristics of strange quark stars within the framework of $f(Q)$ gravity, incorporating the MIT EoS with baryon number density dependent $B$, denoted as $B(n)$. This study presents novel solutions for compact stars within the framework of $f(Q)$ gravity. Our findings reveal several intriguing features of compact stars, demonstrating not only the physical plausibility but also support from recent observations, as demonstrated by the consistency between the observed results and the results predicted from the model. Here, we apply the Finch-Skea metric ansatz to construct a physically viable stellar model. Applying this metric ansatz, we solve the modified EFEs in the presence of anisotropy. It is assumed that interior fluid is composed of deconfined three flavours up, down and strange quarks, encapsulated in a region known as a bag, following the EoS of the MIT bag model for quark matter. In the standard form of MIT EoS, $p_r=\frac{1}{3}(\rho-4B)$, it was assumed that $B$ is a constant quantity and is a measure of the energy difference between perturbative and non-perturbative vacuum. CERN experiments have provided strong evidence of a phase transition from hadronic matter to Quark-Gluon Plasma (QGP) under extreme densities \cite{Witten,Baym,Glendenning}. The MIT bag model, with constant $(B)$, is unable to describe such a phase transition properly. To develop a more realistic and physically viable equation of state, we introduce a baryon number density $(n)$ dependent bag parameter denoted by $B(n)$, inspired by the Wood-Saxon potential \cite{Woods} and parameterised as a function of baryon number density $(n)$.\\
We compute the energy per baryon $(E_B=\rho_0/n)$ as a function of baryon number density $(n)$, and the resulting $(E_B)$ vs $n$ plot is illustrated in Fig.~\ref{figa1}. By analysing the $(E_B)$ vs. $n$ plot, one can find valuable insights into the thermodynamic sustainability of the system, which is essential for understanding and predicting the behaviour under different conditions. It is noted in Fig.~\ref{figa1} that for baryon number density $(n)\leq 0.244~fm^{-3}$, the energy per baryon $(\rho_0/n)$ surpasses $939~MeV$, which represents the typical energy per baryon of a neutron. In this situation, stellar configuration indicates a stable neutron star composed entirely of hadronic matter. The energy per baryon $(\rho_0/n)$ of $^{56}Fe$ is $930.4~MeV$. When $(\rho_0/n)$  lies within the range of $930.4<(\rho_0/n)<939~MeV$, the state is termed as meta-stable state and in this case baryon number density $(n)$ lies within the range of $0.244\leq n \leq 0.247~fm^{-3}$. Within this range of $(\rho_0/n)$, the transition of phase from hadronic matter to quark matter starts. This leads to a meta-stable state, potentially a mixture of both hadronic and quark matter. The formation of a hybrid star is possible under these conditions. Similarly, when $(\rho_0/n)$ lies below the level of $930.4~MeV$, one may expect a stable quark matter configuration. $E_B$ asymptotically approaches a constant value of $26.68~MeV$ for large value of $n=1.4~fm^{-3}$. The reason behind this nature is that, at lower baryon densities, the available energy states within the system are barely populated by quarks. This allows quarks to occupy low-energy states, maximizing their energy per baryon. However, as the baryon density increases, more and more energy states become filled with quarks. This leads to a degenerate state where no additional quarks can occupy low-energy levels. Consequently, the energy per baryon remains constant regardless of further increase in baryon number density $(n)$.
Empirical evidence from nuclear physics indicates that ordinary nuclei do not undergo spontaneous conversion into a hypothetical dense quark phase, consistent with the observed stability of hadronic nuclear matter under ordinary conditions. The strange matter hypothesis originally suggested by Bodmer \cite{Bodmer} and developed by Witten \cite{Witten} proposes that quark matter composed of up, down, and strange quarks could, under some conditions, represent the absolute ground state of strongly interacting matter, with energy per baryon lower than that of normal nuclei \cite{Beiglbock}. However, theoretical treatments using the MIT bag model indicate that non-strange quark matter is not energetically favoured at zero pressure, which is why nuclei are stable and do not spontaneously convert to quark matter \cite{Holdom}. Furthermore, strange quarks $(s)$ are heavier than up $(u)$ and down $(d)$ quarks and baryons containing strangeness are correspondingly more massive than nucleons, making the introduction of strangeness energetically unfavourable in normal nuclear matter. \\
However, the situation is different for quark matter. The Fermi momentum within quark matter, estimated to be between $300$ and $350~MeV$ \cite{Gao}, exceeds the mass of a strange quark. This makes it energetically favourable for some non-strange quarks to transform into strange quarks, reducing the Fermi momentum and overall energy \cite{Witten}, thereby stabilising the system. Thus, for the model, we are considering a stable state of matter composed of three flavours: up $(u)$, down $(d)$ and strange $(s)$ quarks. We also restrict our discussion within the range of $57.55<B(n)<91.55~MeV/fm^{-3}$ for stable strange quark matter. Below the value of $B(n)=57.55~MeV/fm^3$, quark matter would theoretically decay into a system of only up $(u)$ and down $(d)$ quarks. However, this system would not be stable if we assume that the quarks within it do not interact with each other. Now, within the range of $57.55\leq B(n) \leq91.55~MeV/fm^{-3}$, we analyse the model for different values of baryon number density $(n)$.\\
This paper explores the theoretical foundations of the symmetric teleparallel equivalent of GR, a geometric theory of gravity where gravity arises from 'non-metricity' $(Q)$, by considering a linear $f(Q)$ action. Considering the present model, we have analysed the TOV equations to derive the maximum mass that can be supported by the model. We noted that for $n=1.40~fm^{-3}$, maximum mass reaches up to the value of $2.47~M_\odot$. It is noted from Table~\ref{tab:1} that the maximum mass varies significantly with the baryon number density $(n)$. We have analysed the profiles of $\rho,~ p_r,~p_t$ and anisotropy function, $\Delta =p_t -p_r$. We study the causality and energy conditions to establish the physical viability of the model. Furthermore, we also study the stability of the model on the basis of the generalised TOV equation, Herrera cracking condition and the variation of adiabatic index within the star. \\
The graphical representation of energy density $(\rho)$ is shown in Fig.~\ref{fig1}. For physical analysis, we consider the star 4U 1820-30 under $f(Q)$ gravity and observe that the energy density is maximum at the centre and decreases towards the stellar surface. In Fig.~\ref{fig1a}, it is noted that energy density depends on the non-metricity parameter ($\alpha_0$). Energy density is also shifted towards the higher value for a lower value of baryon number density $(n)$ as shown in Fig.~\ref{fig1b}. We also graphically represent the radial pressure $(p_r)$ as well as tangential pressure $(p_t)$, and are shown in Figs.~\ref{fig2} and \ref{fig3} respectively. Within the stellar interior, both radial pressure $(p_r)$ and tangential pressure $(p_t)$ exhibit monotonically decreasing profiles with increasing radial distance from the centre. It is noted from Figs.~\ref{fig2} and \ref{fig3} that $p_r$ and $p_t$ increases with the increase of $\alpha_0$ and baryon number density $(n)$. Dependence of anisotropy $\Delta=(p_t-p_r)$ on the parameters $\alpha_0$ and $n$ is shown in Fig.~\ref{fig4}. The anisotropy in pressure is greatly affected by the non-metricity parameter $(\alpha_0)$. 
The causality condition describes the condition of stability by limiting the velocity of sound inside the stellar configuration. The graphical representation of the causality condition is shown in Figs.~\ref{fig5} and \ref{fig6}. It is noted that the causality condition is well maintained inside the stellar configuration. As we have considered the MIT bag model here, the variation of $v_r^2$ throughout the stellar object is constant, i.e., $\frac{1}{3}$. Its variation is irrespective of the baryon number density $(n)$. Similarly, we have analysed the transverse sound velocity and conclude that $v_t^2$ decreases as we increase the baryon number density $(n)$. The energy conditions are verified through Fig.~\ref{fig1} and Figs.~\ref{fig7a} -\ref{fig7e}. It is also noted that $\alpha_0$ and $n$ can modify the energy conditions. To study the stability of the proposed model, we analyse via the generalised Tolmen-Oppenheimer-Volkoff (TOV) equation. It is graphically shown in Fig.~\ref{fig8}, and we observed that all three forces are balancing each other within the stellar configuration in static equilibrium. Cracking condition proposed by Herrera \cite{Herrera} to analyse the stability of the model under small fluctuations in their physical variables, represented in Fig.~\ref{fig9}, and we observed that it is well maintained in view of the Eq.~(\ref{eq44}). Stability is also checked through the study of adiabatic index $(\Gamma)$ and is shown in Fig.~\ref{fig10}. The radii of several compact stars have been predicted within the framework of the present model and are listed in Table~\ref{tab:2}. It is important to mention that the proposed model, formulated in the context of $f(Q)$ gravity and the MIT EoS with baryon number density dependent bag parameter $B$, allows the determination of radii for several compact stars that are believed to be strange stars. Furthermore, by adopting suitable parametric choices in the $n$-dependent bag model, the radii of several recently observed compact stars can be reproduced with a high degree of accuracy.\\
The preceding analysis suggests that the proposed model provides a physically consistent description of anisotropic stellar configuration composed of three-flavour quark matter in $f(Q)$ gravity theory. Furthermore, the incorporation of a baryon number density-dependent bag parameter $(B)$ yields a more realistic EoS for characterising the internal structure and stability of strange stars. Thus, combining $n$ dependent $B$ and $f(Q)$ gravity formalism, one can establish a physically viable model for strange stars.

\section{Acknowledgments} 
SB and DB are thankful to Department of Physics, Cooch Behar Panchanan Barma University for providing all the necessary help to carry out this research. DB is thankful to the Department of Science and Technology (DST), Govt. of India, for providing the fellowship vide no: DST/INSPIRE Fellowship/2021/IF210761. PKC gratefully acknowledges support from IUCAA, Pune, India under Visiting Associateship programme. This research was funded by Princess Nourah bint Abdulrahman University Researchers Supporting Project number (PNURSP2026R924), Princess Nourah bint Abdulrahman University, Riyadh, Saudi Arabia. 

\end{document}